\documentclass[a4paper,superscriptaddress,twocolumn,pra,showkeys,preprintnumbers,floatfix,showpacs]{revtex4-1}

\usepackage{epsfig}
\usepackage{amsmath}
\usepackage{amssymb}
\usepackage{grffile}
\usepackage{color}
\usepackage[dvipsnames]{xcolor}
\usepackage{ulem}

\newcommand{\ket}[1]{\left |#1\right \rangle}

\newcommand{\expec}[1]{\left \langle #1\right \rangle}

\newcommand{\mc}{\mathcal}

\newcommand{\beq}{\begin{eqnarray*}}
\newcommand{\eeq}{\end{eqnarray*}}

\newcommand{\be}{\begin{eqnarray}}
\newcommand{\ee}{\end{eqnarray}}

\usepackage[colorlinks,bookmarks=false,citecolor=blue,linkcolor=red,urlcolor=blue]{hyperref}

\begin{document}

\title{Correlations and enlarged superconducting phase of $t$-$J_\perp$ chains of ultracold molecules on optical lattices}

\author{Salvatore R. Manmana}
\affiliation{Institute for Theoretical Physics, Universit\"at G\"ottingen, Friedrich-Hund-Platz 1, D-37077 G\"ottingen, Germany}
\author{Marcel M\"oller}
\affiliation{4th Physical Institute, Universit\"at G\"ottingen, Friedrich-Hund-Platz 1, D-37077 G\"ottingen, Germany}
\author{Riccardo Gezzi}
\affiliation{Institute for Theoretical Physics, Universit\"at G\"ottingen, Friedrich-Hund-Platz 1, D-37077 G\"ottingen, Germany}
\author{Kaden R.~A. Hazzard}
\affiliation{Department of Physics and Astronomy, Rice University, Houston, Texas 77005, USA and Rice Center for Quantum Materials, Rice University, Houston, Texas 77005, USA}
\preprint{NSF-KITP-15-166}

%\pacs{CHECK,71.10.Fd, 71.10.Hf, 71.10.Pm,74.20.Mn}
\date{\today}

\begin{abstract}
We compute physical properties across the phase diagram of the $t$-$J_\perp$ chain with long-range dipolar interactions, which 
describe ultracold polar molecules on optical lattices. 
Our results obtained by the density matrix renormalization group (DMRG)
indicate that 
superconductivity is enhanced when the Ising component $J_z$ of the spin-spin interaction and the charge component $V$ are tuned to zero,
and even further by the long-range dipolar interactions.
At low densities,  a substantially larger spin gap is obtained.
We provide evidence that long-range interactions lead to algebraically decaying correlation functions despite the presence of a gap.
Although this has recently been observed in other long-range interacting spin and fermion models, the correlations in our case have the peculiar property of having a small and continuously varying exponent.
We construct simple analytic models and arguments to understand the most salient features.
\end{abstract}

\maketitle

\section{Introduction}
\label{sec:intro}

At the interface between atomic, molecular, optical, and condensed matter physics, systems of ultracold polar  molecules~\cite{reviewmolecules,Lemeshko_Review,PhysRevLett.101.133004,Silke_science,Silke_science2,Silke_Nature,
PhysRevLett.104.030402,
Amodsen,PhysRevLett.84.246,PhysRevLett.94.203001,PhysRevLett.101.133005,deMiranda:2011gd,reviewmolecules,
PhysRevLett.112.070404,focus_ultracoldmolecules,PhysRevLett.114.205302,
PhysRevLett.116.225306,review_molecules2,MolonyPRL,TakekoshiPRL,GoulvenPRL2016} have caused a great deal of excitement and opened a path for the quantum simulation~\cite{Manin,feynman1,feynman2,feynman3,MolPhys2013,micheli_naturephysics,pupillo_polargases,PhysRevB.87.081106,KadenPRL2014} of quantum magnetism~\cite{springer_quantummagnetism,book_frustratedspins,book_HFMTrieste,Simon:2011p2830,1367-2630-9-5-138,reviewmolecules_quantummagnetism,
PhysRevLett.110.075301,Yan:2013fn,PhysRevA.84.033619} and superconductivity~\cite{dagotto,PhysRevLett.107.115301} on optical lattices~\cite{Bloch:2005p988,Bloch:2008p943}. 
Intrinsic to these systems are the long-range dipolar-type interactions which, in contrast to the long-range Coulomb interaction in condensed matter systems, are not affected by screening. 
As described in Refs.~\onlinecite{PhysRevLett.107.115301,PhysRevA.84.033619}, these systems can emulate generalized $t$-$J$-type models~\cite{tJ1977,tJoriginal1,tJoriginal2,auerbach,dagotto} in which the values of all parameters can be tuned independently by forming dressed states of the molecules via DC electrical fields and microwaves. 
We consider the experimentally simplest variant,
in which the spin exchange is anisotropic and of the $XX$ type, which we call the $t$-$J_\perp$ model. 
As discussed in Ref.~\onlinecite{PhysRevA.84.033619}, this model arises in a parameter regime accessible to experiments with reactive ultra-cold polar molecules; indeed, experiments out-of-equilibrium in this parameter regime have observed spin-exchange interactions~\cite{PhysRevLett.110.075301,Yan:2013fn}.   
Furthermore, the same model can describe nonreactive molecules in a realistic parameter regime, even when accounting for the complicated collisional physics~\cite{MaylePRA2012,MaylePRA2013},
as follows from the results in Refs.~\onlinecite{AndrisPRL2016,MichaelWallPRA2017,MichaelWallPRA2017_2}.

\begin{figure}[t]
\includegraphics[width=0.4115\textwidth]{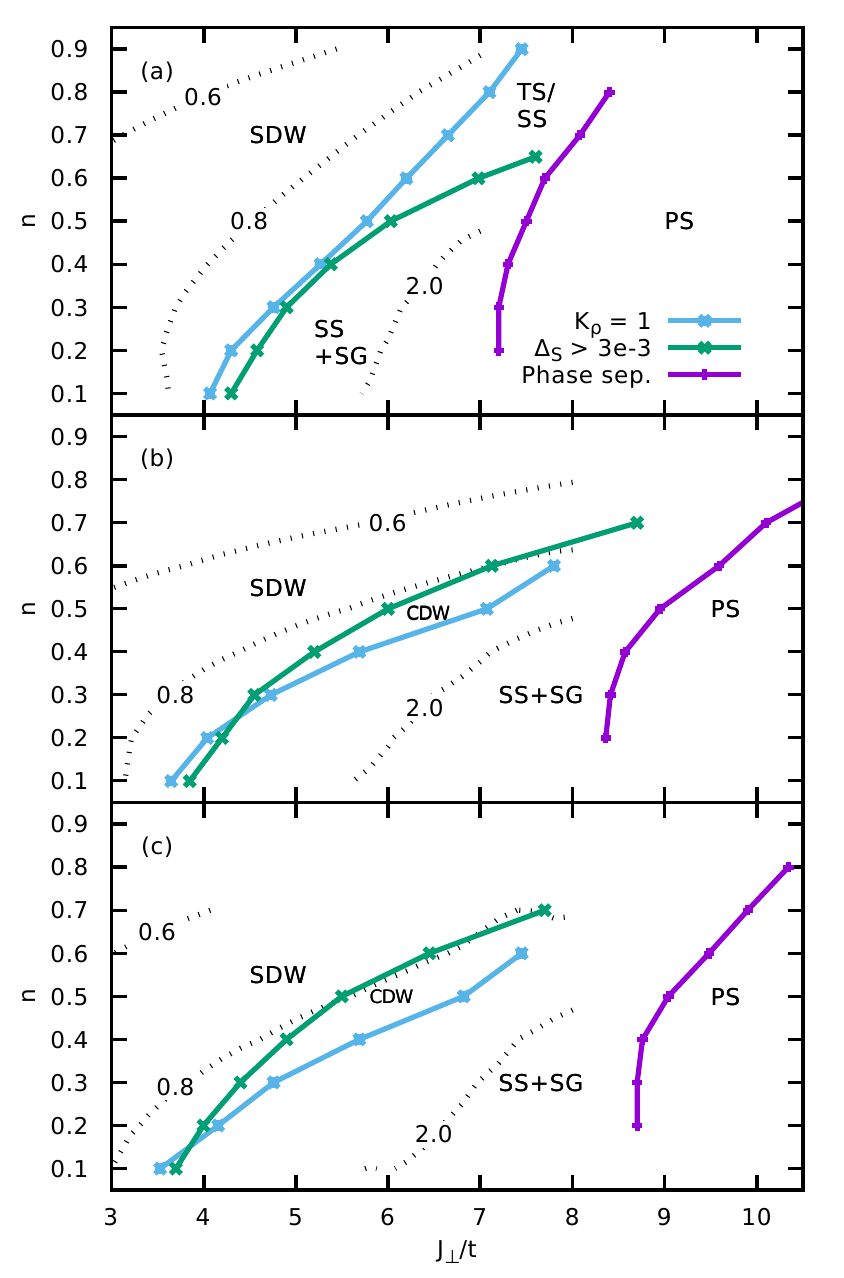}
\caption{Phase transition and crossover lines obtained by the DMRG after extrapolating to the thermodynamic limit: (a) $t$-$J_\perp$ chain with NN interactions~\eqref{eq:nntJperp} (also see Ref.~\cite{PhysRevLett.107.115301}); (b) $t$-$J_\perp$ model with NNN interactions~\eqref{eq:nnntJperp}; (c) dipolar $t$-$J_\perp$ model~\eqref{eq:dipolartJperp}. 
The different regions are SDW, the two-channel Luttinger liquid (LL) with dominant spin-density wave (SDW) correlations; SS+SG, the singlet superconductor (SS) with finite spin gap~\eqref{eq:spingap} (SG), a one-channel LL phase; PS, phase separation, characterized by a vanishing inverse compressibility~\eqref{eq:compressibility}; and CDW, the one-channel LL phase with dominating charge-density wave (CDW) correlations and finite spin gap. The dotted lines indicate constant values of the Luttinger parameter $K_\rho$ and the blue line indicates $K_\rho = 1$. At the green line, the spin gap becomes larger than $3 \times 10^{-3}t$ (estimated numerical accuracy after extrapolating to the thermodynamic limit) upon increasing $J_\perp/t$. The purple line indicates the onset of phase separation.
We display only results for $J_\perp/t > 3$, since for smaller values the systems seem to be in the SDW phase. 
Note that in (b) and (c) we present numerical results for the phase transition and crossover lines only for densities $n \leq 0.7$, as described in the text. 
Also note that in (a) we do not further distinguish between triplet SC (TS) and SS, since we are neglecting logarithmic corrections, which can make the TS channel dominant~\cite{giamarchi,voitreview}.
}
\label{fig:phasedia}
\end{figure}

In this paper, we address the phase diagram of the one-dimensional $t$-$J_\perp$-model with anisotropic spin interactions in the $XX$ limit and in which the density-density interaction is set to zero.
We start with nearest-neighbor (NN) interactions~\cite{PhysRevLett.107.115301} and then extend the study to systems with next-nearest-neighbor (NNN) and long-range dipolar-type interactions as realized in polar molecules.  
We apply the density-matrix renormalization group (DMRG)~\cite{white1992,white1993,dmrgbook,noack:93,Schollwock:2005p2117,Schollwock:2011p2122}.
Figure~\ref{fig:phasedia} shows the phase diagrams we obtained.  
While for NN interactions we are able to obtain the phase diagram for fillings $0.1 \leq n \leq 0.9$, for the NNN and the dipolar case it is difficult to obtain reliable results at fillings $n \gtrsim 0.7$, so for these cases we focus on the behavior at $n \lesssim 0.7$.
On a qualitative level, the resulting phase diagrams are very similar to the one of the usual SU(2)-invariant $t$-$J$ chain with NN interactions (see Ref.~\onlinecite{Moreno} and references therein) with the main difference being an additional intermediate phase with finite spin gap and Luttinger parameter~\cite{giamarchi} $K_\rho < 1$ appearing in the case of NNN and dipolar interactions.
This is similar to what was obtained in Ref.~\onlinecite{Troyer1993} when adding density-density interactions to the $t$-$J$ model. 
On the quantitative level, when using energy units in which the hopping amplitude is set to one, $t \equiv 1$, the extensions of the metallic and superconducting phases and the magnitude of the spin gap are enhanced. 
More subtly, we find that in the presence of the dipolar interactions an {\it algebraic} tail is visible in the long-distance decay of correlation functions even in the gapped phase. 
A similar effect has been reported previously for Ising-systems with dipolar interactions~\cite{DengPRA2005,SchachenmayerNJP2010}, for long-range interacting $p$-wave superconductors with Majorana edge modes~\cite{AlexeyGuidoPRL2014}, and for systems with quadratic  algebraically decaying interactions it can be proven that in such systems the correlation functions decay with the same exponent as the one of the interaction~\cite{SchuchCommMathPhys2006}.
Here, we report such an effect in an intriguing regime. 
Similar to the aforementioned results, we have a spin gap with algebraically decaying correlations. 
However, the current system has at least two distinct features: First, there is an additional, gapless (charge) degree of freedom, and, second, we find a correlation that decays spatially much more rapidly than the interaction itself, with an exponent that varies continuously with the model parameters.

The paper is organized as follows.
In Sec.~\ref{sec:models} we introduce the models, methods, and observables used to analyze the phase diagrams shown in Fig.~\ref{fig:phasedia}.
In Sec.~\ref{sec:phasedias} we present in some detail the numerical results that we used to derive the phase diagrams of Fig.~\ref{fig:phasedia}. 
In Sec.~\ref{sec:variational} we develop a toy model to estimate the size of the spin-gapped superconducting region at low fillings and provide a simple understanding for why this phase gets enhanced when one adds an anisotropy to the spin interactions or tunes the density-density interactions. 
In Sec.~\ref{sec:longrange} we discuss the effect of long-range interactions and provide perturbative arguments for how they lead to an algebraic tail of the correlation functions in a gapped phase.  
In Sec.~\ref{sec:summary}, we summarize.
We restrict in this paper to purely one-dimensional systems and work in units $\hbar \equiv 1$.

\section{Models, Methods, Observables, and phase diagram}
\label{sec:models}
\subsection{Models}
\label{subsec:models}

Polar molecules in optical lattices are described by \cite{PhysRevLett.107.115301,PhysRevA.84.033619} 
\begin{equation}
\begin{split}
&\mathcal H^{tJVW} =  - t \sum_{i,\sigma}  \left[c^\dagger_{i,\sigma} c^{\phantom{\dagger}}_{i+1,\sigma} + \textrm{h.c.}\right] \\
&+ \sum_{j>i} \frac{1}{|i-j|^3} \left[  \frac{J_\perp}{2} \left( S^+_i S^-_j  + S^-_i S^+_j \right) + J_z S^z_{i} S^z_{j} \right. \\
& \left. \phantom{\frac{J_\perp}{2}} + V n^{\phantom{\dagger}}_i n^{\phantom{\dagger}}_j + W n^{\phantom{z}}_i S^z_{j}  \right],   
\end{split}
\label{eq:tjvw}
\end{equation}
where we assume that double occupancies are not allowed. 
As usual, $c^{(\dagger)}_{i,\sigma}$ are fermionic annihilation (creation) operators for a particle with spin $\sigma$ on lattice site $i$, the Hilbert space is the usual fermionic Hilbert space projected onto the space with no doublons (as in the usual $t$-$J$ model), 
$S^+_i = c^\dagger_{i,\uparrow}c^{\phantom{\dagger}}_{i,\downarrow}$ and $S^-_i = c^\dagger_{i,\downarrow}c^{\phantom{\dagger}}_{i,\uparrow}$ are the spin raising and lowering operators, $S^z_i = (c^\dagger_{i,\uparrow} c^{\phantom{\dagger}}_{i,\uparrow} - c^\dagger_{i,\downarrow} c^{\phantom{\dagger}}_{i,\downarrow} )/2$ is the $z$ component of the spin operator, and $n^{\phantom{\dagger}}_i = \sum_{\sigma} c^\dagger_{i,\sigma} c^{\phantom{\dagger}}_{i,\sigma}$ is the total density on site $i$.  
Note that all parameters of the Hamiltonian as well as the filling can be tuned independently.
Remarkably, in the polar molecule realization it is also possible to emulate a bosonic version of this Hamiltonian; here, however, we restrict ourselves to the fermionic case due to its relation to condensed-matter systems. 
Also note that in the cold-molecule implementation the hopping term is only between nearest-neighbor lattice sites, while the other terms (originating from the dipolar interaction between the molecules) are long ranged.  
In this paper we treat systems in one spatial dimension.

References~\onlinecite{PhysRevLett.107.115301,PhysRevA.84.033619} showed that Eq.~\eqref{eq:tjvw} describes molecules in a sufficiently deep optical lattice that the doublons are suppressed. 
This suppression could result either from strong Hubbard-type on-site interactions or from quantum Zeno suppression of the doublons in reactive molecules. 
Recently, it has been argued theoretically that even for nonreactive molecules, the on-site interactions are not Hubbard-like and involve numerous interaction channels~\cite{AndrisPRL2016,MichaelWallPRA2017,MichaelWallPRA2017_2}.
Despite this, when the multichannel interactions are sufficiently strong,  doublons are suppressed and Eq.~\eqref{eq:tjvw} is the correct effective description of the system. 

The model~\eqref{eq:tjvw} is a generalization of the standard $t$-$J$ model \cite{tJoriginal1,tJoriginal2,auerbach,dagotto}, which in one dimension (1D) reads 
\begin{equation}
\begin{split}
\mathcal H ^{tJ} = - t \sum_{i,\sigma}& \left[ c_{i,\sigma}^{\dagger} c_{i+1,\sigma}^{\phantom{\dagger}} + \textrm{h.c.} \right] \\
& + J \sum_i \left[ \vec{S}^{\phantom{\dagger}}_i \cdot \vec{S}^{\phantom{\dagger}}_{i+1} - \frac{1}{4} n^{\phantom{\dagger}}_i n^{\phantom{\dagger}}_{i+1} \right], 
\end{split}
\label{eq:standard_tJ}
\end{equation}
and which is obtained via second-order degenerate perturbation theory from the Hubbard model~\cite{auerbach}, therefore retaining the SU(2) symmetry of the original model. 
In perturbation theory, one finds $J = 4t^2/U$, with $U$ the strength of the Hubbard interaction, and it is not possible to tune the parameters $t,\, J_\perp, \, J_z,$ and $V$ independently from each other. 
Note that model~\eqref{eq:standard_tJ} is obtained from Eq.~\eqref{eq:tjvw} by considering only nearest neighbor interactions and setting $J_z = J_\perp \equiv J$ and $V = -J/4$.

Although the molecular system is in principle fully tunable, the simplest experimental realization of model~\eqref{eq:tjvw} has $J_z = V = W = 0$~\cite{PhysRevLett.107.115301,PhysRevA.84.033619,Yan:2013fn}, motivating us to calculate the phase diagram of the dipolar $t$-$J_\perp$ chain
\begin{equation}
\begin{split}
\mathcal H^{tJ_\perp} = &- t \sum_{i,\sigma} \left[ c_{i,\sigma}^{\dagger} c_{i+1,\sigma}^{\phantom{\dagger}} + \textrm{h.c.} \right] \\
&+ \frac{J_\perp}{2} \sum_{j>i} \frac{1}{|i-j|^3} \left[ S_i^+ S_j^- + S_i^- S_j^+ \right]  \, .
\end{split}
\label{eq:dipolartJperp}
\end{equation}
We also consider the model truncated to only NN interactions,
\begin{equation}
\begin{split}
\mathcal H^{{\rm NN} \, tJ_\perp} = &- t \sum_{i,\sigma} \left[ c_{i,\sigma}^{\dagger} c_{i+1,\sigma}^{\phantom{\dagger}} + \textrm{h.c.} \right] \\
&+ \frac{J_\perp}{2} \sum_{i} \left[ S_i^+ S_{i+1}^- + S_i^- S_{i+1}^+ \right] \, ,  
\end{split}
\label{eq:nntJperp}
\end{equation}
and NNN interactions, 
\begin{equation}
\begin{split}
&\mathcal H^{{\rm NNN}\, tJ_\perp} = - t \sum_{i,\sigma} \left[ c_{i,\sigma}^{\dagger} c_{i+1,\sigma}^{\phantom{\dagger}} + \textrm{h.c.} \right] \\
&+ \frac{J_\perp}{2} \sum_i \left[ \left( S_i^+ S_{i+1}^- + S_i^- S_{i+1}^+ \right) + \frac{1}{8} \left( S_i^+ S_{i+2}^- + S_i^- S_{i+2}^+ \right) \right] \, .   
\end{split}
\label{eq:nnntJperp}
\end{equation}
Comparing the dipolar case to these models with truncated interaction ranges isolates the effects of the long-range interaction. 

As can be seen in Fig.~\ref{fig:phasedia}(a), the ground-state phase diagram of the $t$-$J_\perp$ chain with NN spin exchange is similar to the one of the standard $t$-$J$ chain~\eqref{eq:standard_tJ}~\cite{Moreno}: The phases and the overall structure of the phase diagram are unchanged, but the numerical values of the phase boundaries are modified.
This indicates that competing interactions that govern the central physics of the $t$-$J$ chain are
taken into account by the interplay of the kinetic energy term with the $J_\perp$ term. 
Note that Ref.~\onlinecite{Troyer1993} found that adding density-density interactions (both NN and long-range) influences the size of the superconducting region in the phase diagram.
Also note that Ref.~\onlinecite{BatistaOrtizPRLtJz} obtained the phase diagram of the $t$-$J_z$-chain by an exact Bethe \textit{Ansatz} calculation. 
It also shows the sequence of metallic, superconducting, and phase separation phases, but in particular at low fillings it differs qualitatively from the phase diagram of the standard $t$-$J$ model, so the behavior of the $t$-$J_\perp$ model appears to be closer to the one of the $SU(2)$ symmetric model. 
These studies, however, show that both the $V$-term and the $J_z$-term can cause (or at least influence) the superconducting (SC) phase. 

As can be seen in Fig.~\ref{fig:phasedia}, setting $J_z=0$ and $V=0$ in 1D strongly enhances the superconducting phase in comparison to the result of Ref.~\onlinecite{Moreno}.
This finding raises the question of what the optimal $V$ and $J_z$ values are for superconductivity.  
In Sec.~\ref{sec:variational} we will address the interplay of the $J_\perp$-, the $J_z$-, and the $V$-term with the kinetic energy at low fillings, which gives us excellent estimates for the phase boundaries and allows us to discuss the importance of each of these terms for the SC phase.

Turning to the effect of dipolar interactions, we note that Ref.~\onlinecite{PhysRevLett.107.115301} discussed the possibility of a phase that was absent for the NN~model, where there is a spin gap but $K_\rho<1$.
Due to the long-range interactions, the numerical treatment is more difficult so that 
Ref.~\onlinecite{PhysRevLett.107.115301} was unable to reach a definitive conclusion.
Here, we extend these calculations to treat larger system sizes and confirm the presence of such an intermediate phase, which we identify as a charge-density wave (CDW). 
In order to do so, we analyze in Sec.~\ref{sec:longrange} the effect of keeping the interaction terms at all distances.
We find that the intermediate phase does seem to persist in the numerics. 
We also find that a cutoff in the interaction range of the order of 10-20 sites can lead to excellent quantitative agreement of the observables treated, so  it becomes possible to obtain reliable results for larger systems.
Interestingly, this works best for gapless phases. 
As soon as a gap opens, the long-range nature of the interactions leads to an algebraic tail that dominates the usual exponentially decaying correlation function, as discussed in detail in Sec.~\ref{sec:longrange}. 
\subsection{Observables}
\label{subsec:observables}
To obtain the phase diagrams shown in Fig.~\ref{fig:phasedia}, we follow Ref.~\onlinecite{Moreno} and analyze a variety of observables described in this section.

An important indicator for the expected Luther-Emery-like phase is the spin gap
\begin{equation}
\Delta_S = E_0(N,S^z_{\rm total} = 1) - E_0(N,S^z_{\rm total} = 0) \, .
\label{eq:spingap}
\end{equation}
This spin-gapped superconducting phase is expected to phase separate at larger $J_\perp/t$, which is characterized by a vanishing inverse compressibility
\begin{eqnarray}
\kappa ^{-1} (n)  & = & n^2 \frac{\partial ^2 e_0(n)}{\partial n^2}
\nonumber \\ & \approx &
 n^2 \frac{[e_0(n+\Delta n) + e_0(n-\Delta n) - 2e_0(n)]}{\Delta n^2},
\label{eq:compressibility}
\end{eqnarray}
where $e_0(n)$ denotes the ground-state energy per site at filling $n$. 

We note that our system is not invariant under SU(2) spin transformations and so we must use the appropriate bosonization expressions for correlation  functions accounting for this. 
In general, one requires a dressed charge matrix~\cite{book_hubbardmodel} to describe this situation. 
In such a case, the spin and charge degrees of freedom mix into new effective degrees of freedom. 
However, in the present case, as there is no external field, U(1) symmetry remains and we can apply the SU(2) bosonization expressions, with the only modification that $K_\sigma \ne 1$ in general.

The Luttinger parameter $K_\rho$ can be related to the Fourier transform of the density-density correlation functions
\begin{equation}
N_{ij} = \langle n_i n_j \rangle - \langle n_i \rangle \langle n_j \rangle,
\label{e1}
\end{equation}
which reads
\begin{equation}
N(k) = \frac{1}{L} \sum _{i,j=1} ^L e^{ik(i-j)} N_{ij}.
\label{eq:Nk}
\end{equation}
From bosonization, it is known that for a gapless Luttinger liquid (LL) phase (i.e., no finite spin or charge gap) the density correlation function is~\cite{Voit} (we do not consider logarithmic corrections)
\begin{eqnarray}
\langle n(r)n(0) \rangle  & = & \frac{K_\rho}{(\pi  r)^2}
+ A_1 \frac{\cos(2k_F r)}{r^{K_\sigma+K_\rho}} %\ln^{-3/2} (r)
\nonumber \\ &  &
+ A_2 \cos(4k_F r) r^{-4K_\rho} 
\label{eq:ncorr}
\end{eqnarray}
and in a phase with finite spin gap
\begin{eqnarray}
\langle n(r)n(0) \rangle  & = &  \frac{K_\rho}{(\pi  r)^2} + A_1 \cos(2k_F r)r^{-K_\rho}.
\label{ncorr-gap}
\end{eqnarray}
Hence, as long as the charge gap is zero, $K_\rho$ can be obtained from the limit $k \rightarrow 0$ of $N(k)$,
\begin{equation}
N(k) \rightarrow K_\rho \frac{|k|}{\pi} \quad  \text{ for } k \rightarrow 0
\label{e6}
\end{equation}
by fitting the slope of $N(k)$.
In a finite system of length $L$, one fits $N(k)$ over a range $1/L <  k < 1/\ell$, where $\ell$ is the relevant microscopic length; one does this for several $L$ and extrapolates to the thermodynamic limit.
Similarly, to obtain $K_\sigma$, we can use the Fourier transform of spin correlation functions \cite{Voit} 
\begin{equation}
\langle S^\alpha(r) S^\alpha(0) \rangle   =  \frac{K_\sigma}{(\pi  r)^2}
+ A \frac{\cos(2k_F r)}{r^{\gamma_{\rm SDW}}} \; , %\ln^{-3/2} (r)
%\nonumber \\ &  & + A_2 \cos(4k_F r) r^{-4K_\rho} \; ,
\label{eq:scorr}
\end{equation}
where for the spin components $\alpha = x,y$, the exponent $\gamma_{\rm SDW, x} = \gamma_{\rm SDW, y} = K_\rho + K_\sigma^{-1}$, and for $\alpha = z$, $\gamma_{\rm SDW, z} = K_\rho + K_\sigma$. 
As above for $K_\rho$, the numerical value of $K_\sigma$ can hence be obtained, e.g., by computing the longitudinal spin correlation function
\begin{equation}
C^{\rm spin, long}_{ij} = \langle S_i^z S_j^z \rangle  - \langle S_i^z \rangle \langle S_j^z \rangle
\label{eq:SzSz}
\end{equation}
and by fitting the slope of the spin structure factor 
\begin{equation}
\begin{split}
S(k) =  & \frac{1}{L} \sum _{i,j=1} ^L e^{ik(i-j)} C^{\rm spin, long}_{ij} \\
& \rightarrow K_\sigma \frac{|k|}{\pi} \quad \text{ for } k \rightarrow 0 \;.
\end{split}
\label{eq:Ksigma}
\end{equation}
However, extrapolation and interpretation of the  numerical results requires care. 
At $k=0$, the bosonization expressions~\eqref{eq:ncorr},~\eqref{ncorr-gap}, and~\eqref{eq:scorr} lead to a value of the structure factor in the thermodynamic limit, which is exactly zero.
However, we find that for finite systems as $k\rightarrow 0$, $N(k)$ and $S(k)$ asymptote not to zero, but to a small finite value, which for a fixed value of $L$ gets larger when approaching the phase-separation region. 
This could indicate that the bosonization expressions might lose their validity in this region.
However, we attribute this to finite-size effects, which become more pronounced close to phase separation. 
Indeed, over a wide range of parameters, the value of the structure factors at $k=0$ seems to go to zero when performing a finite-size extrapolation so that in the thermodynamic limit again $N(k=0)=0$ and $S(k=0)=0$. 
We therefore believe our results to build a valid basis for obtaining the numerical values of $K_\rho$ and $K_\sigma$ also in the case of long-range interactions.
Nevertheless, the finite-size effect is more pronounced for long-range interactions.
At large fillings ($n \gtrsim 0.7$), this makes it difficult to investigate for the existence of a superconducting precursor region to phase separation. 
The possible SC phase is expected to be small and the finite-size effects in the region of interest are so substantial that the value of the structure factors at $k=0$ for the system sizes treated could not safely be extrapolated to zero, so fitting values for $K_\rho$ becomes meaningless. 
Alternatives would be to treat much larger systems or to perform the computations directly in $k$ space, as, e.g., discussed in Ref.~\onlinecite{ejima05}.
Since both would be a substantial numerical effort, we leave such studies to future research.

In the following we use the behavior of $K_\rho$, $K_\sigma$, the compressibility, and the spin gap to determine the phase diagrams. 
We complement this by analyzing the algebraic decay of various correlation functions and identifying the dominant ones (i.e., the ones that decay with the smallest exponent). 
In particular, we consider the density-density correlation function~\eqref{e1}, the  longitudinal spin correlation function~\eqref{eq:SzSz}, the transverse spin-spin correlation function, which is independent of the longitudinal correlations since the $t$-$J_\perp$ model lacks SU(2) invariance, 
\begin{equation}
C^{\rm spin, trans}_{ij} = \langle S_i^+ S_j^- \rangle,
\label{eq:Strans}
\end{equation}
and the pairing correlation functions
\begin{equation}
P_{ij} = \langle \Delta^{\dagger}_{T,S}(i) \, \Delta^{\phantom \dagger}_{T,S}(j) \rangle.
\label{e3}
\end{equation}
Here
\begin{equation}
\Delta^{\dagger}_S(i) = \frac{1}{\sqrt{2}} \left(c_{i,\downarrow} ^\dagger c_{i+1,\uparrow} ^\dagger - c_{i,\uparrow} ^\dagger c_{i+1,\downarrow} ^\dagger \right)
\label{e3.1}
\end{equation}
for singlet pairing and
\begin{equation}
\Delta^{\dagger}_T(i) = c_{i,\uparrow} ^\dagger c_{i+1,\uparrow} ^\dagger
\label{e3.2}
\end{equation}
for triplet pairing. 
Note that the lack of SU(2) invariance imposes up to three different possible triplet pairing channels; for simplicity, we focus only on the one defined in Eq.~\eqref{e3.2}. 
In a gapless LL, bosonization predicts~\cite{pruschke92,voitreview}
\begin{equation}
\begin{split}
\langle  \Delta _S ^\dagger (r) &\Delta _S (0) \rangle   =\\
&  C_0 r^{-(K_\sigma+1/K_\rho)} + C_1 \cos(2k_F r)r^{-(K_\rho+1/K_\rho)}
\end{split}
\label{PScorr}
\end{equation}
and in the presence of a spin gap ($K_\sigma$ is not defined anymore)
\begin{equation}
\begin{split} 
\langle  \Delta _S ^\dagger (r) & \Delta _S (0) \rangle   =\\ 
& C_0' r^{-1/K_\rho} + C_1' \cos(2k_F r) r^{-(K_\rho +1/K_\rho)} \, ,
\end{split} 
\label{PScorr-gap}
\end{equation}
while for the triplet pairing correlations in the absence of a spin gap 
\begin{equation}
\begin{split} 
\langle  \Delta _T ^\dagger (r) & \Delta _T (0) \rangle   = C_0'' r^{-(1/K_\sigma + 1/K_\rho)} +\\
& C_1'' \cos(2k_F r)r^{-(K_\rho+1/K_\rho + K_\sigma + 1/K_\sigma)}
\end{split} 
\label{PTcorr}
\end{equation}
and in the presence of a spin gap 
\begin{equation}
\langle  \Delta _T ^\dagger (r) \Delta _T (0) \rangle   =  C_0''' r^{-1/K_\rho}
+ C_1''' \cos(2k_F r) r^{-(K_\rho +1/K_\rho)} \, .
\label{PTcorr-gap}
\end{equation}
Motivated by the various bosonization results for the correlation functions listed in this section, we fit our numerical results for each of the correlations to a function of the form
\begin{equation}
\begin{split}
f(|i-j|) = &\frac{A}{|i-j|^\alpha} + \\
&B \frac{\cos(k_1 |i-j| + \varphi_1)}{|i-j|^\beta} + C \frac{\cos(k_2 |i-j| + \varphi_2)}{|i-j|^\gamma}
\end{split}
\label{eq:fit}
\end{equation}
in order to obtain the values of the exponents and also of $K_{\rho, \sigma}$ from a direct fit to the real space data. 
This complements the momentum space fits of the structure factors described earlier.
We expect $\varphi_1=\varphi_2=0$ and, as a function of filling $n$, the Fermi wave vector is $2k_F = n \pi$. 
We expect $k_1 = 2 k_F$ and $k_2 = 4 k_F$.
However, we obtain more stable fits by fitting these variables.
By comparing the so obtained values for $\alpha, \, \beta,$ and $\gamma$ of the different correlation functions, we identify the dominant one by choosing the smallest absolute values of these fitting parameters.  
Note that due to the complicated fitting function, the error in the values of the exponents can be $\sim 20\%$ (see, e.g., Ref.~\onlinecite{BLBQ} for a similar study in a spin system) and sometimes even larger.  
We have performed the fits by a direct least-squares fitting procedure and also using more powerful genetic algorithms \cite{CRAN_GA}.
The results of both approaches are of comparable quality. 
In the following we present the best obtained fit results, typically from the simpler direct least-squares fitting procedure. 

\subsection{DMRG}
\label{subsec:DMRG}

We apply the DMRG method~\cite{white1992,white1993,dmrgbook,noack:93,Schollwock:2005p2117,
Schollwock:2011p2122} for obtaining ground-state properties of the system in the presence of NN, NNN, and long-range dipolar-type interactions.
In all cases, we apply open boundary conditions (OBC) to systems with up to $L=200$ lattice sites. 
For systems with long-range interactions, we keep track of all interaction terms when $L \leq 100$.
For larger systems, we introduce a cutoff for distances $d>20$, which due to the smallness of the interaction beyond this distance usually is in  quantitative agreement with taking the full range of interactions into account, as demonstrated in Sec.~\ref{sec:longrange} in the case of gapless phases.
For the relevance of the long-range interactions on the behavior of correlation functions in the presence of a gap, see Sec.~\ref{sec:longrange}. 
Typically, we perform 6-10 sweeps and keep up to $m=1000$ states. 
Despite the presence of long-range interactions, the calculations over a wide range of the phase diagram appear to be well converged and the results for obtaining the phase diagram have an accuracy comparable to the ones typically obtained for NN interactions. 
Convergence problems mainly arise in the vicinity of the phase-separation region and at high densities, where we are particularly careful when discussing the results. 
Note that it appears useful for future studies to use a formulation of the algorithm in terms of matrix product operators, as, e.g., discussed in Refs.~\onlinecite{PhysRevB.87.081106,PaeckelSciPost} and references therein, which speeds up the calculations. 

\section{Luttinger parameter, Spin Gap, Compressibility, and Exponents of the Correlation Functions}
\label{sec:phasedias}

In this section we discuss how we obtained the phase diagrams presented in Fig.~\ref{fig:phasedia} by using the observables and fitting procedures discussed in Sec.~\ref{subsec:observables}. 

\subsection{The NN $t$-$J_\perp$ chain}

\begin{figure}
\includegraphics{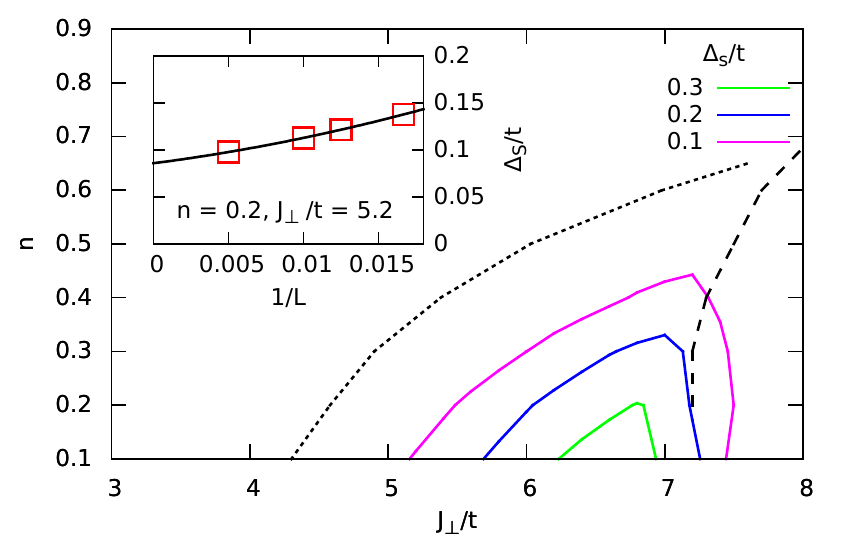}
\caption{Spin gap~\eqref{eq:spingap} as a function of filling $n$ and $J_\perp/t$ for the NN model~\eqref{eq:nntJperp}. The short-dashed line indicates $\Delta_S/t = 3 \times 10^{-3}$, which was used in Fig.~\ref{fig:phasedia} as the border of the spin-gapped region. The long-dashed line indicates the onset of the phase separation region, as in Fig.~\ref{fig:phasedia}. The inset shows typical finite-size scaling behavior.}
\label{fig:DeltaSnn}
\end{figure}

Here we revisit the phase diagram of the $t$-$J_\perp$ chain of Ref.~\onlinecite{PhysRevLett.107.115301} and discuss in more detail some of its features.
We start with the spin gap. 
An example for a typical finite-size extrapolation using a quadratic fit function is shown in the inset of Fig.~\ref{fig:DeltaSnn}, leading to the spin gap in the thermodynamic limit displayed in the main panel of Fig.~\ref{fig:DeltaSnn}.
A spin-gapped region, i.e., a Luther-Emery liquid, is readily apparent.
We identify the boundary of this region by the contour line on which the gap is $3 \times 10^{-3}t$, which we estimate to be the accuracy of our finite-size extrapolation. (The extrapolated data start to show artifacts at smaller values. Note that, e.g., in Ref.~\onlinecite{MichaudPRB2010} by comparison to the Bethe \textit{Ansatz} for an $XXZ$ chain an even higher accuracy of $5 \times 10^{-4}$ was estimated. For the long-range interactions, however, the convergence is more difficult to control, so we use this more conservative error estimate.)  
Outside this region we assume the gap to be zero, or at least so small that it cannot be  resolved reliably. 
Interestingly, the spin gap appears to decrease close to the region where phase separation occurs (diagnosed by the compressibility, as discussed below), resulting in a pronounced maximum around $J_\perp/t\sim 6.6$.

\begin{figure}[b]
\includegraphics{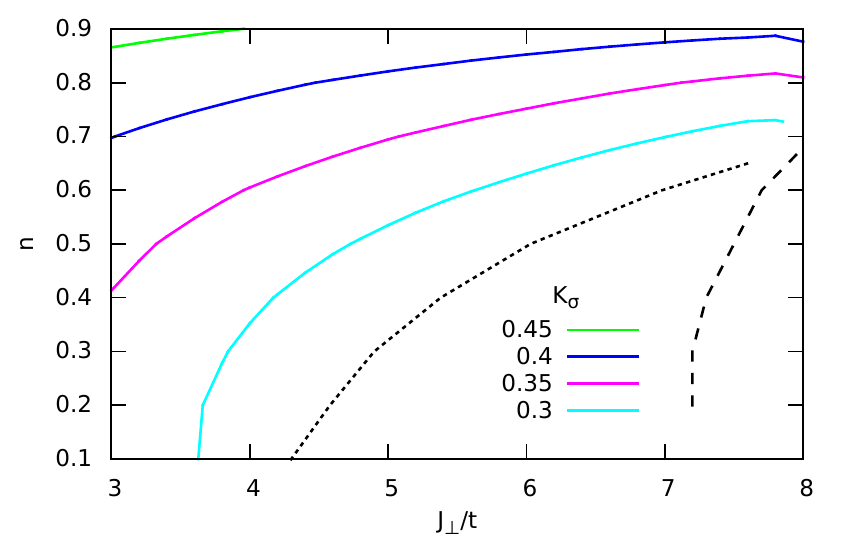}
\caption{Luttinger liquid parameter $K_\sigma$ for the NN model~\eqref{eq:nntJperp} as a function of the filling $n$ and of $J_\perp/t$ obtained from fitting the structure factor of the longitudinal spin correlations~\eqref{eq:Ksigma} as discussed in the text. 
The short-dashed line indicates $\Delta_S/t = 3 \times 10^{-3}$, which was used in Fig.~\ref{fig:phasedia} as the border of the spin gapped region. The long-dashed line indicates the onset of phase separation, as in Fig.~\ref{fig:phasedia}. 
}
\label{fig:Ksigma_nn}
\end{figure}

Next we consider the values of the Luttinger parameters $K_\rho$, displayed in Fig.~\ref{fig:phasedia}(a), and $K_\sigma$, displayed in Fig.~\ref{fig:Ksigma_nn}.  
In Fig.~\ref{fig:phasedia} the region with $K_\rho <1$ can be distinguished from the one with $K_\rho>1$, in which pairing correlation functions become dominant \cite{giamarchi}.
In Fig.~\ref{fig:Ksigma_nn} we show our results for $K_\sigma$ as a function of $J_\perp/t$ and $n$.

\begin{figure}
\includegraphics{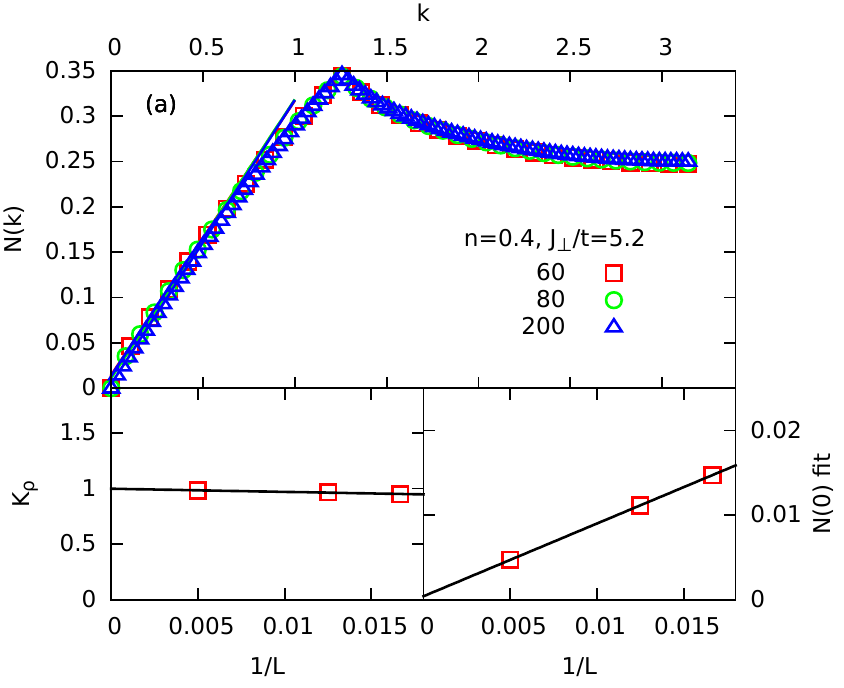}
\includegraphics{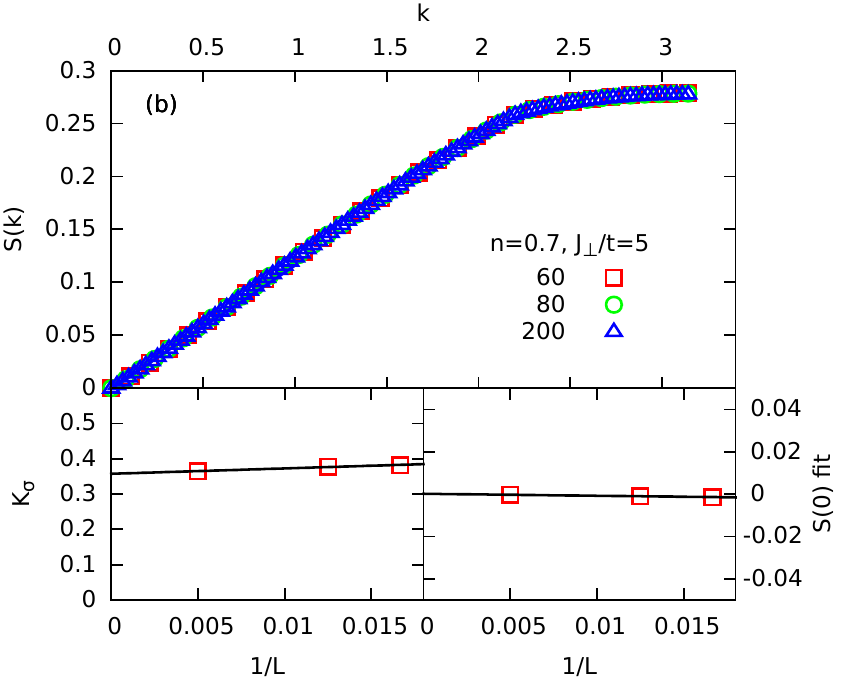}
\caption{Typical fits for obtaining the Luttinger liquid parameters of the NN model~\eqref{eq:nntJperp} from the structure factors of the respective correlation functions. (a) Fit of the slope for $k \to 0$ in (a) the charge structure factor from  Eq.~\eqref{eq:Nk} and (b) the structure factor of the longitudinal spin correlation function~\eqref{eq:Ksigma}. The insets show the extrapolation to the thermodynamic limit: left panel, $K_{\rho,\sigma}$; right panel, value of $N(k)$ or $S(k)$ at $k=0$.}
\label{fig:LLparamsextrapolnn}
\end{figure}

\begin{figure}
\includegraphics{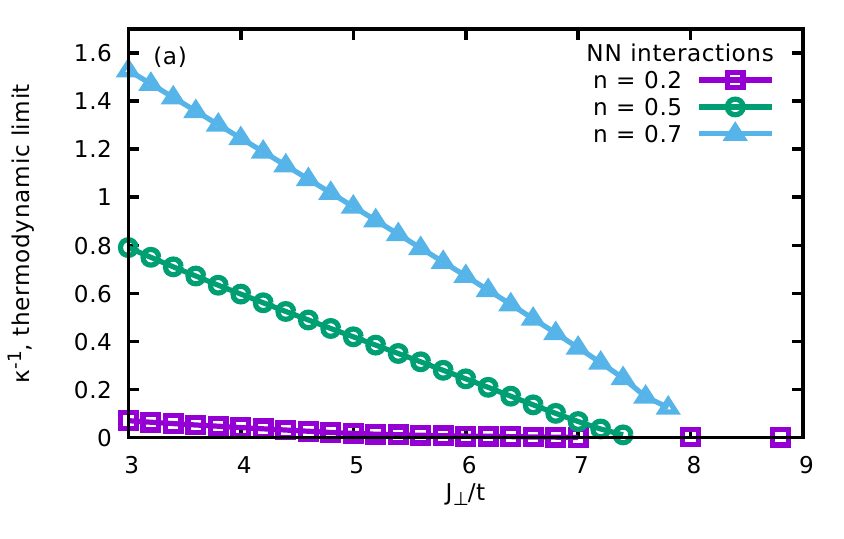}
\includegraphics{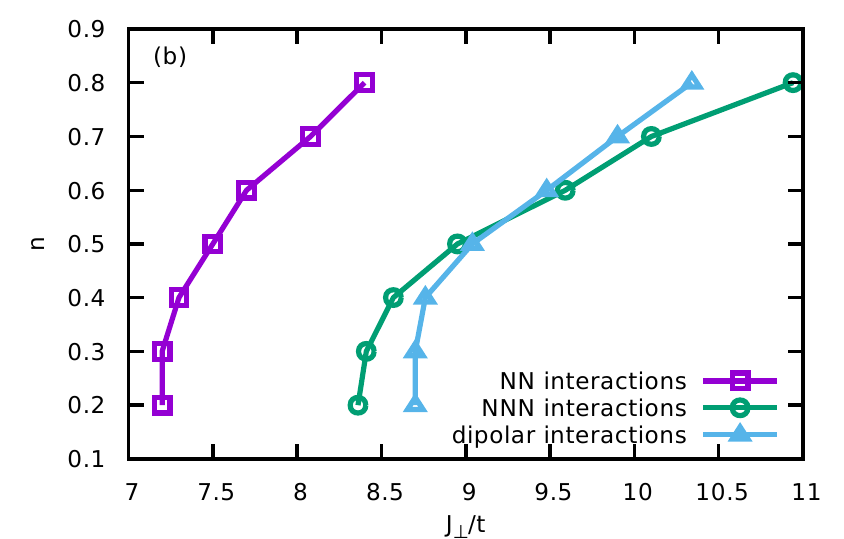}
\caption{(a) Inverse compressibility $\kappa^{-1}$ [Eq.~\eqref{eq:compressibility}] for different values of the filling $n$ as a function of $J_\perp/t$ for the system with nearest-neighbor interactions~\eqref{eq:nntJperp}. (b) Line in the $(n, J_\perp)$ plane at  which $\kappa^{-1} = 0$, indicative of  the phase separation region for the chain with NN [Eq.~\eqref{eq:nntJperp}, purple squares], NNN [Eq.~\eqref{eq:nnntJperp}, green circles], and dipolar [Eq.~\eqref{eq:dipolartJperp}, blue triangles] interactions.}
\label{fig:inversecompressibilitynn}
\end{figure}

\begin{figure}[b]
\includegraphics{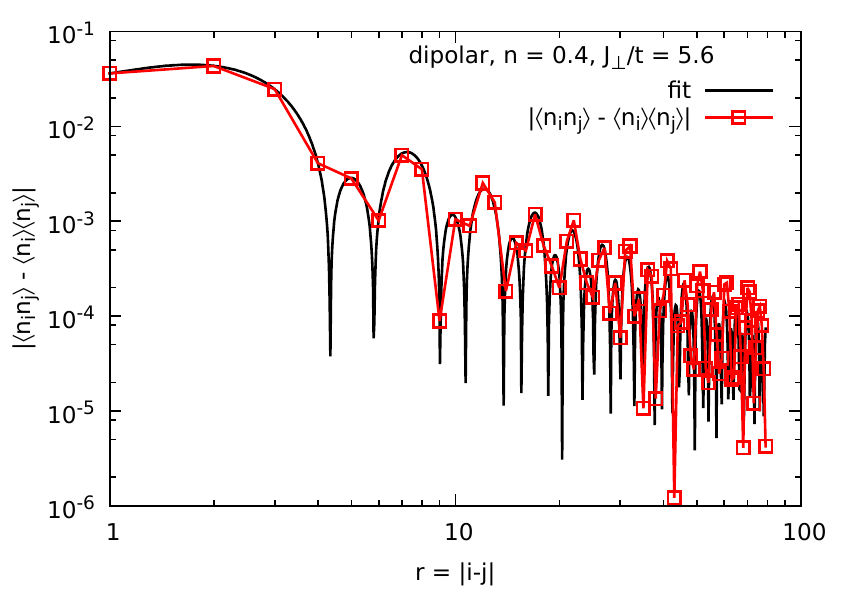}
\caption{Example of a fit of the density-density correlation function~\eqref{e1} using Eq.~\eqref{eq:fit}.}
\label{fig:exponentsnn}
\end{figure}

In Fig.~\ref{fig:LLparamsextrapolnn}(a) we show a typical example of how we obtained the values of $K_\rho$ by fitting the slope in the charge structure factor for results on finite systems and extrapolating to the thermodynamic limit and in Fig.~\ref{fig:LLparamsextrapolnn}(b) the same procedure for obtaining $K_\sigma$ from fitting to the spin structure factor.
For $J_\perp/t \gtrsim 8$ it becomes more difficult to keep the high numerical accuracy, making it more difficult to control the obtained values for $K_\rho$. 
That this region coincides with the region in which phase separation is obtained is displayed in Fig.~\ref{fig:inversecompressibilitynn}, showing the inverse compressibility going to zero in this region.
For completeness, we also performed real space fits of the correlation functions to the LL expressions.
In Fig.~\ref{fig:exponentsnn} we show an example of the fits using Eq.~\eqref{eq:fit}. 

\subsection{The dipolar and NNN $t$-$J_\perp$ chains}

Now we perform the same analysis for the dipolar and NNN variants of the $t$-$J_\perp$ chain, Eqs.~\eqref{eq:dipolartJperp} and~\eqref{eq:nnntJperp}, respectively.

In both cases, the spin gap behaves very similarly to the one in the NN model, but 
it is larger, a point to which we will return in Sec.~\ref{sec:variational}. 
Its value for the NNN system is larger than in the NN case and is further increased in the case of dipolar interactions.
Also, taking into account the longer-range interactions leads to a further extension of the spin-gapped region towards smaller as well as larger values of $J_\perp/t$, which we will also address in Sec.~\ref{sec:variational}. 

\begin{figure}
\includegraphics{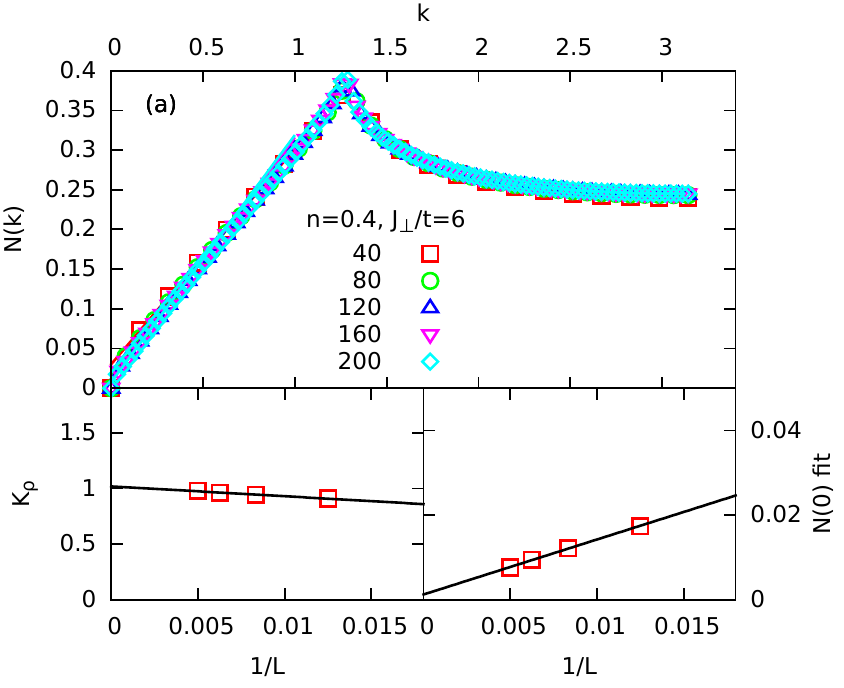}
\includegraphics{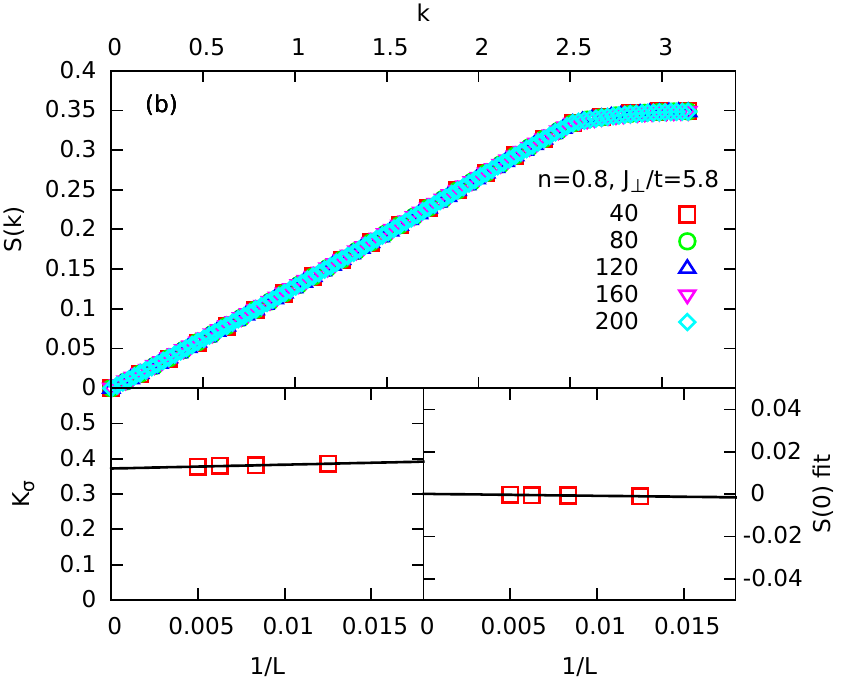}
\caption{Typical fits of the structure factors to obtain (a) $K_\rho$ and (b) $K_\sigma$ for the system with dipolar interactions. The results look similar for the NNN case. The insets show extrapolations to the thermodynamic limit of the values of the Luttinger liquid parameters as well as of the values of the structure factors at $k=0$.}
\label{fig:Kextrapol_dipolar}
\end{figure}

Next we consider the Luttinger parameters $K_\rho$ and $K_\sigma$.
The values we obtained for $K_\rho$ are displayed in Figs.~\ref{fig:phasedia}(b) and~\ref{fig:phasedia}(c) for the systems with NNN and dipolar interactions, respectively. 
Again, they were obtained by fitting the $k \to 0$ part of the charge structure factor as in Eq.~\eqref{e6}, for which we show typical examples in Fig.~\ref{fig:Kextrapol_dipolar}. 
Note that we perform the fits assuming an $A+ b k$ behavior of the structure factors at $k=0$, where $A$ disappears in the thermodynamic limit as shown in the insets of Fig.~\ref{fig:Kextrapol_dipolar}, as discussed in Sec.~\ref{subsec:observables}. 
However, in these cases, at large fillings when approaching phase separation, it becomes more and more difficult to perform this analysis, as the value of $N(k)$ for $k \to 0$ becomes larger with increasing filling and $J_\perp/t$. 
It is difficult to tell if this is due to convergence problems, which are not apparent from the calculations, or if other effects come into play.
As we base our analysis of the phase diagram on the behavior of $K_\rho$ and obtaining it becomes more difficult in this region, we discuss the phase diagram only up to fillings $n = 0.7$.
It appears necessary to consider more elaborate tools to extract the value of $K_\rho$ more reliably in this region, e.g., by computing the structure factor directly in momentum space \cite{ejima05}.  
Similar problems are encountered when computing $K_\sigma$, so we refrain from discussing its behavior for the NNN and dipolar systems.

Note that, in contrast to the NN case, the regions with a finite spin gap and with $K_\rho < 1$ now overlap in a small intermediate region of the phase diagram, as indicated in Figs.~\ref{fig:phasedia}(b) and~\ref{fig:phasedia}(c).
The appearance of this intermediate phase is the main qualitative difference of the phase diagram caused by considering spin-exchange interactions beyond nearest neighbors. 

\begin{figure}
\includegraphics{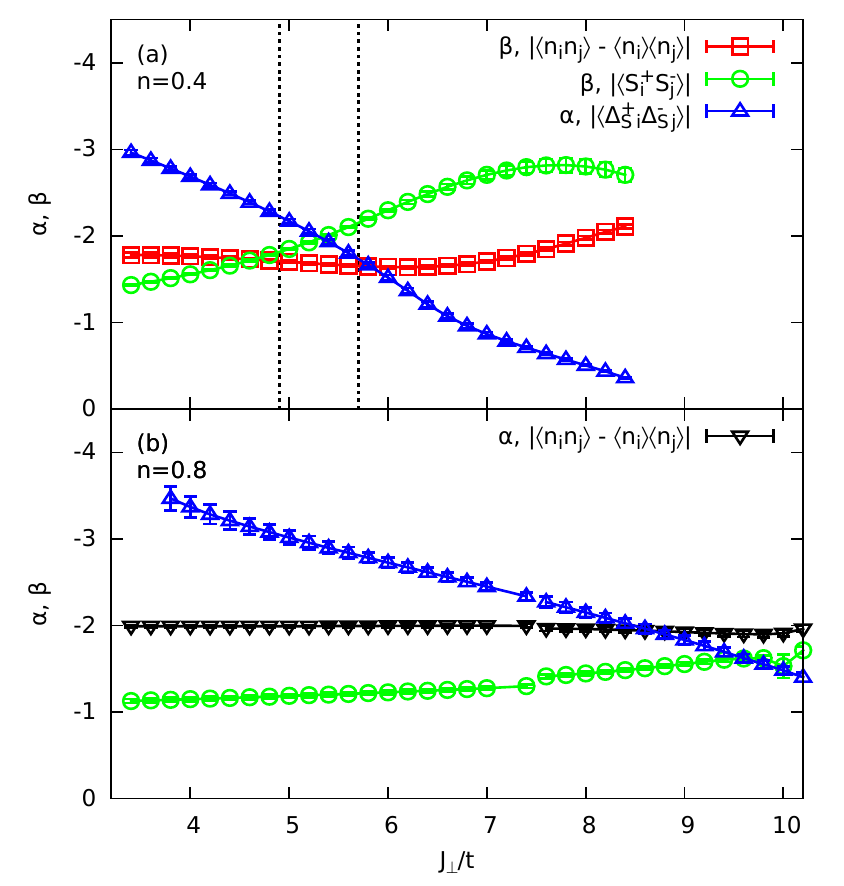}
\caption{Typical results for the exponents of the correlation functions of the system with dipolar interactions~\eqref{eq:dipolartJperp}: the dominant exponents are shown at filling (a) $n = 0.4$ and (b) $n=0.8$ as a function of $J_\perp/t$. The dotted vertical lines in (a) indicate the position of the intermediate CDW phase obtained from the spin gap and the $K_\rho=1$ line in Fig.~\ref{fig:phasedia}.}
\label{fig:exponents_dipolar}
\end{figure}

In order to further characterize the phases in detail, we turn to the behavior of the exponents of the correlation functions, 
which we obtain by fitting Eq.~\eqref{eq:fit} to the numerical results in real space.
This gives us an independent estimate for the phase boundaries.
In Fig.~\ref{fig:exponents_dipolar} we show the three correlations with the smallest value of the exponent in the dipolar case, from which we identify the dominant correlation at long range. 
Figure~\ref{fig:exponents_dipolar}(a) shows the results at $n=0.4$.
As can be seen, an intermediate region appears in which CDW correlations become dominant.  
The boundaries of this region are in good quantitative agreement with the results of Fig.~\ref{fig:phasedia} based on $K_\rho$ and the opening of the spin gap.  
At $n=0.8$ [Fig.~\ref{fig:exponents_dipolar}(b)], the results indicate that at $J_\perp/t \gtrsim 9.6$ the singlet-pairing correlations become dominant, which would support the presence of a small SC precursor region to phase separation also at large densities.
To confirm the presence of such a phase, it would be necessary to obtain $K_{\rho,\sigma}$ with high accuracy, as mentioned before. 

Summarizing, our results for the various observables strongly concur for the phase diagrams shown in Fig.~\ref{fig:phasedia}.
Questions that arise at this point are why the spin-gapped phase is enhanced by setting to zero the $J_z$ and $V$ terms and furthermore by the long-range interactions, as indicated by the comparison at low filling $n=0.1$ displayed in Fig.~\ref{fig:spingap_n0.1}. 
Also, it remains to clarify if the dipolar interactions may have a further effect on the long-distance behavior of correlation functions.
Both are further discussed in the following two sections. 

\section{Understanding the phase diagram: variational approach}
\label{sec:variational}

In this section we present an analytical approach to understanding why at low fillings there are two transitions and how their locations depend on the model parameters.
Similar considerations can be found in Refs.~\onlinecite{Troyer1993,Dagotto1992,Kivelson1990}.
Here we adapt them to the situation of fully tunable parameters as realizable in the polar molecule quantum simulators.  
Although we make use of a rather crude variational approach applied to a toy model, we end up with good predictions of all of the trends and of the right order of magnitude for the values of the critical points. 

\begin{figure}[b]
\includegraphics[width=0.48\textwidth]{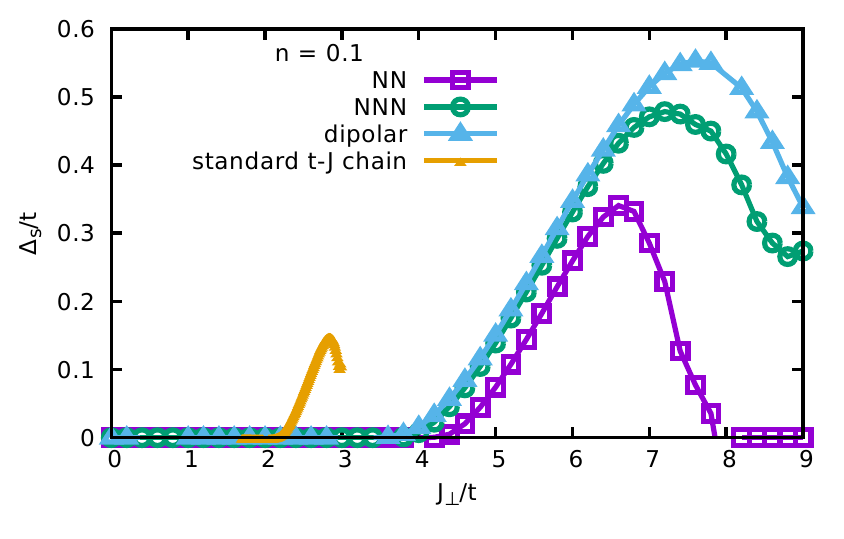}
\caption{Spin gap $\Delta_S/t$ at filling $n=0.1$ for the usual $t$-$J$ chain [Eq.~\eqref{eq:standard_tJ}, data from Ref.~\onlinecite{Moreno}], the $t$-$J_\perp$ chain with NN interactions [Eq.~\eqref{eq:nntJperp}], with NNN interactions [Eq.~\eqref{eq:nnntJperp}], and with dipolar interactions [Eq.~\eqref{eq:dipolartJperp}] as a function of $J_\perp/t$ ($J/t$ for the standard $t$-$J$ chain).}
\label{fig:spingap_n0.1}
\end{figure}

The DMRG results of Fig.~\ref{fig:phasedia} and of Ref.~\onlinecite{Moreno} show that for both $t$-$J$ and $t$-$J_\perp$ Hamiltonians, at low fillings a superconducting phase forms in a window of values of $J$ or $J_\perp$ for a broad range of densities $n$. 
Namely, upon increasing $J$ or $J_\perp$, the system is driven from a gapless  Luttinger liquid into the superconducting spin-gap phase at a critical value $J_c^{(1)}$. 
Then there is a transition from the superconducting phase to a phase-separated region at $J_c^{(2)}$. 
The DMRG shows several interesting trends for the coupling constants $J_c$.
First, both values of $J_c$ increase with density. 
Next, the values of both $J_c$'s are larger for the $t$-$J_\perp$ case than for the standard $t$-$J$ case. 
Additionally, the superconducting regime is wider for the $t$-$J_\perp$ case.
This is doubly important because this wider region also naturally leads to a larger maximum spin gap at low fillings (in units of $t$) as shown in Fig.~\ref{fig:spingap_n0.1}.
Finally, if one moves from nearest-neighbor to dipolar interactions, then the superconducting region widens again: $J_c^{(1)}$ decreases and $J_c^{(2)}$ increases. 
As before, the wider superconducting region gives rise to an increased spin gap.
Our analytical approach reproduces all of these features. 

\subsection{Estimates of the energy of each phase}

The basic idea behind our estimates is to compare cartoons of the three phases captured by two or three particles, and to work in the dilute limit $n\ll1$ to make the estimates simple. 
In particular, we will consider the following three states and models of them:
%\begin{itemize}
%\item 
a spin-gapped phase (superconductor), where we consider the energy of a singlet on two nearest-neighbor sites and a third particle far away;
%\item 
a phase-separated state, with three particles on adjacent sites; 
%  \item 
and a Luttinger liquid, with three far-away particles.
%\end{itemize}
Each state consists of three particles to simplify comparisons of their energy.

In the third case, each particle sits at the bottom of the band $-2t$, for a total energy of $-6t$. The next two sections provide energy estimates for the first two states, and comparing these will let us qualitatively understand the phase diagram.

\subsubsection{Spin gap phase, not phase separated}

Here we consider the energy of a state with two adjacent particles (which, due to the antiferromagnetic coupling in our case, form a singlet) and a far away mobile particle; as a first, simple estimate we calculate the energy in the limit $J_\perp/t$ is large. 
The mobile particle has an energy contribution $-2t$, while the singlet has energy 
$-0.5 t$ for the $J_\perp$ interaction and $-0.75 t$ for the usual $J$ interaction.
Motion of the singlet occurs through higher-order terms, such as $t^2/J_\perp$, and thus can be neglected in the limiting case. However, the real physical case where the transitions into this state occur are not deep in this limit and the singlet motion can be relevant. This is an important point that we return to shortly. 

Since in the following we want to discuss the role of the $J_\perp$, $J_z$, and $V$ interactions, we will express the energies in terms of these three couplings.  
We so obtain for the energy of a singlet plus an independent particle  
\be
E_{SG} &=& - \frac{J_\perp}{2} - \frac{J_z}{4} + V -2t \\
&\equiv&  - \frac{J_\perp}{2} - \frac{\alpha J_\perp}{4} + \beta J_\perp -2t,
\label{eq:sg}
\ee
where we introduced the parametrization $J_z = \alpha J_\perp$ and $V = \beta J_\perp$. 
Further contributions arise, e.g., when the mobile particle is close to the singlet there is an additional interaction, but in the $n\ll 1$ region without phase separation, this happens with negligible frequency.  

For long-range interactions, the singlet can delocalize a bit while still benefiting from the magnetic lowering of the energy, but roughly this can be incorporated by small modifications of $\alpha$ and $\beta$.

\subsubsection{Phase separated region}

A philosophy similar to the one of the preceding section estimates the energy of the phase separated state by considering three adjacent particles. 
We can ignore all tunneling terms if $t\ll J_\perp$.
Now, in the magnetization sector with a single spin up, we simply diagonalize the Hamiltonian in the basis $(\ket{\uparrow\downarrow\downarrow},\ket{\downarrow\uparrow\downarrow},\ket{\downarrow\downarrow\uparrow})$, i.e.,  
the Hamiltonian matrix 
\be
\begin{pmatrix}
2V          & -J^\perp/2  & 0 \\
-J^\perp/2 & 2V-J^z/2      & -J^\perp/2 \\
0          & -J^\perp/2  &    2V  
\end{pmatrix}
\, .
\ee
This has a ground-state energy of
\be 
E_{PS} &=& 2V - \frac{1}{4} \left( \sqrt{8 J_\perp^2 + J_z^2} + J_z \right)\\
&=& J_\perp \left[ 2 \beta - \frac{1}{4} \left( \sqrt{8 + \alpha^2} + \alpha \right) \right]
\ee
with the above parametrization for $J_z$ and $V$. 
 
\subsection{Phase diagram from variational estimates}

To determine the phase diagram at low fillings, we compare the energies of the spin gap and phase-separated state to each other and to the state with three far away mobile particles, which has energy $-6t$. 

\subsubsection{LL-SG transition: $J_c^{(1)}$}

\begin{figure}[t]
\includegraphics{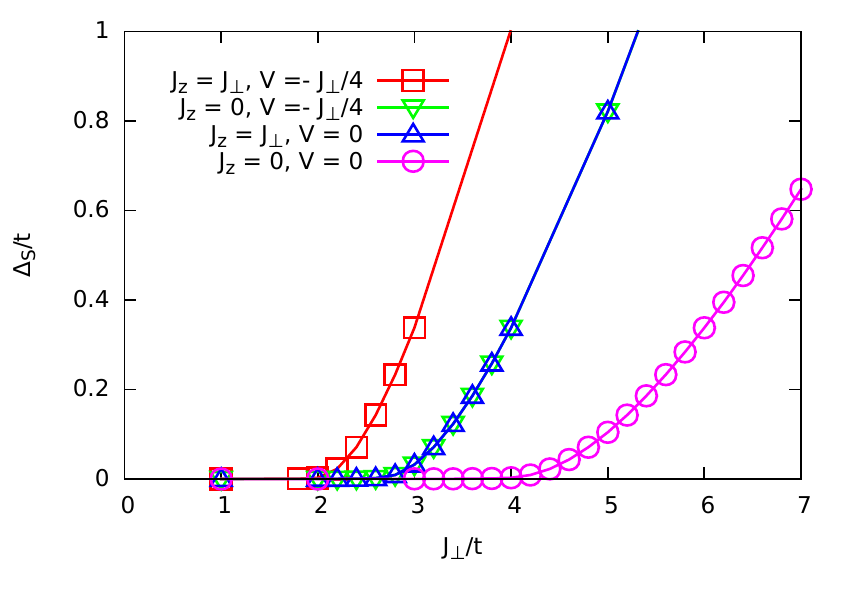}
\caption{The DMRG results for the spin gap~\eqref{eq:spingap} for two particles on a lattice with $L=100$ sites for systems with NN interactions:  red squares: standard $t$-$J$ model~\eqref{eq:standard_tJ}; green down triangles, chain with $J_z=0$ but $V=-J_\perp/4$; blue up triangles, $V=0$ but $J_z = J_\perp$; and magenta circles, $J_z = V = 0$ as in the NN $t$-$J_\perp$ chain~\eqref{eq:nntJperp}.}
\label{fig:spingap2particles}
\end{figure}

The critical point $J_c^{(1)}$ is found by setting $E_{SG}$ equal to the energy of three free particles, i.e., 
\be
J_\perp \left( \beta - \frac{1}{2} - \frac{\alpha}{4} \right) - 2t &=& -6t
\ee
giving
\be
 J_c^{(1)} &=& \frac{16t}{\alpha - 4 \beta + 2}. 
\label{eq:Jc1}
\ee

For $\beta = 0$, this gives $J_c^{(1)}/t = 8$ in the $J_\perp$ case with $\alpha = 0$ and $J_c^{(1)}/t = 16/3t \approx 5.4t$ for SU(2) spin interactions with $\alpha = 1$.
For the standard $t$-$J$ case with $\beta=-1/4$, this results in $J_c^{(1)} = 2t$ in the SU(2) case and $J_c^{(1)} = 8/3t$ in the $J_\perp$ case. 
The latter exactly coincides with the $V=0$, SU(2) case.
This is an interesting observation that we will come return to.  
In Fig.~\ref{fig:spingap2particles} we show the spin gap calculated numerically with the DMRG for a system of two particles on a lattice with 100 sites as a function of $J_\perp/t$ for the different cases. The values of $J_c^{(1)}$ at which the gap opens are in qualitative agreement with our estimate, as well as with the low-density phase transition points in Fig.~\ref{fig:phasedia}. 
Thus, going from the SU(2) case to the $J_\perp$ case shifts $J_c^{(1)}$ to larger values, as one observes from the DMRG results, and the roles of $J_z$ and an interaction $V$ are interchangeable. 

Going from nearest-neighbor to dipolar interactions increases the energy contribution by $J_\perp, \, J_z,$ and $V$, thereby decreasing $J_c^{(1)}$, again in agreement with DMRG calculations.

In fact, the quantitative agreement for this transition is dramatically improved by a simple phenomenological treatment of the nonperturbative singlet motion. Simply allowing a kinetic energy contribution $-2t$ for singlet motion modifies the energy of the spin-gapped state from Eq.~\eqref{eq:sg} to
\begin{equation}
E_{SG} = - \frac{J_\perp}{2} - \frac{\alpha J_\perp}{4} + \beta J_\perp -4t.
\label{eq:sg-ph}
\end{equation}
Then 
\be
 J_c^{(1)} &=& \frac{8t}{\alpha - 4 \beta + 2}. 
 %(2t+V)/\alpha.
\label{eq:Jc1-ph}
\ee
For $\beta=0$, this gives $J_c^{(1)}=4$ in the $J_\perp$ case with $\alpha=0$ and $J_c^{(1)}/t=8/3t\approx 2.7t$ for SU(2) spin interactions with $\alpha=1$. For the standard $t$-$J$ case 
with $\beta=-1/4$ this results in $J_c^{(1)}=2t$ in the SU(2) 
case and $J_c^{(1)}=8/3t$ in the $J_\perp$ case. This is in excellent agreement with the phase diagram we found in
 Fig.~\ref{fig:phasedia}, Ref.~\onlinecite{Moreno}, and the results of Fig.~\ref{fig:spingap2particles}. It is remarkable that such a simple approximation captures the highly nonperturbative physics of the singlet motion.

\subsubsection{SG-PS transition: $J_c^{(2)}$}

The critical point $J_c^{(2)}$ is determined by $E_{SG}=E_{PS}$, and using the above results, we obtain
\be
%J_c^{(2)} &=& (2t + V)/(\beta-\alpha). 
J_c^{(2)} &=& \frac{8t}{ \sqrt{8 + \alpha^2} -4 \beta - 2 } \, .
\label{eq:Jc2}
\ee

For $V=0$ this gives $J_c^{(2)}/t \approx 9.7$ for $\alpha = 0$ and $J_c^{(2)}/t=8$ for the usual SU(2) case with $\alpha=1$.
For $V=-J_\perp/4$ as in the standard $t$-$J$ model, this yields $J_c^{(2)}/t \approx 4.4$ for the $\alpha = 0$ case and $J_c^{(2)}/t = 4$ in the SU(2) case.  
Again, the values are comparable to the DMRG results for systems with NN interactions at low fillings, though systematically larger: At filling $n=0.1$, the DMRG finds for $V=0$ and $\alpha = 0$ a value  $J_c^{(2)}/t \approx 7$, and in Ref.~\onlinecite{Moreno} for the standard $t$-$J$ model  $J_c^{(2)}/t \approx 3$ is reported.
Note that this deviation probably is caused by completely neglecting the kinetic energy in the phase-separation region, which would lower the energy and so lead to a smaller value of $J_c^{(2)}$.
Nevertheless, this approximation is useful for its simplicity and because it captures the main features of $J_c^{(2)}$.
  
Three observations can be made. 
First, in all cases, this value is larger than $J_c^{(1)}$, explaining the existence of two transitions and their ordering (i.e., phase separation does not happen before the spin gap forms). 
Next, going from SU(2) to $J_\perp$ shifts $J_c^{(2)}$ to larger values, as observed by the DMRG. 
Finally, the width $J_c^{(2)}-J_c^{(1)}$ increases going from SU(2) to $J_\perp$, which also increases the maximum spin gap.
All of these features are consistent with the DMRG.

\begin{figure}
\includegraphics[width=0.45\textwidth]{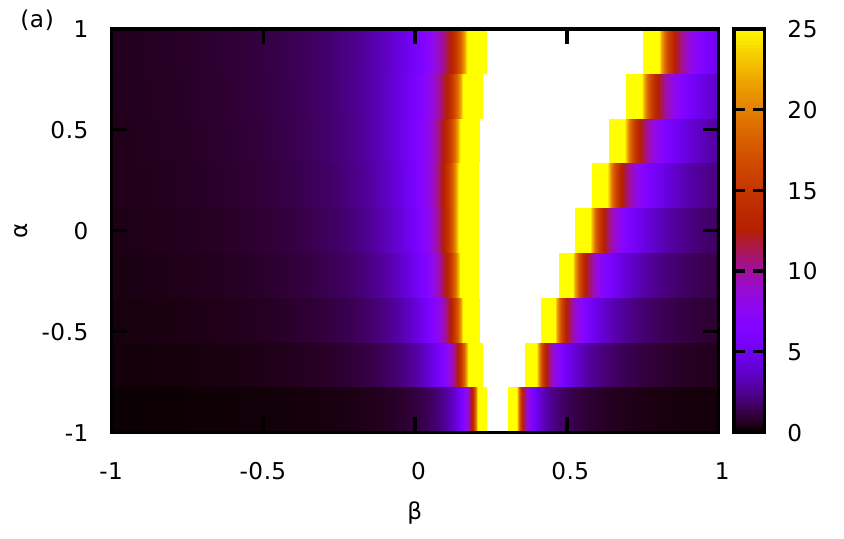}
\includegraphics[width=0.45\textwidth]{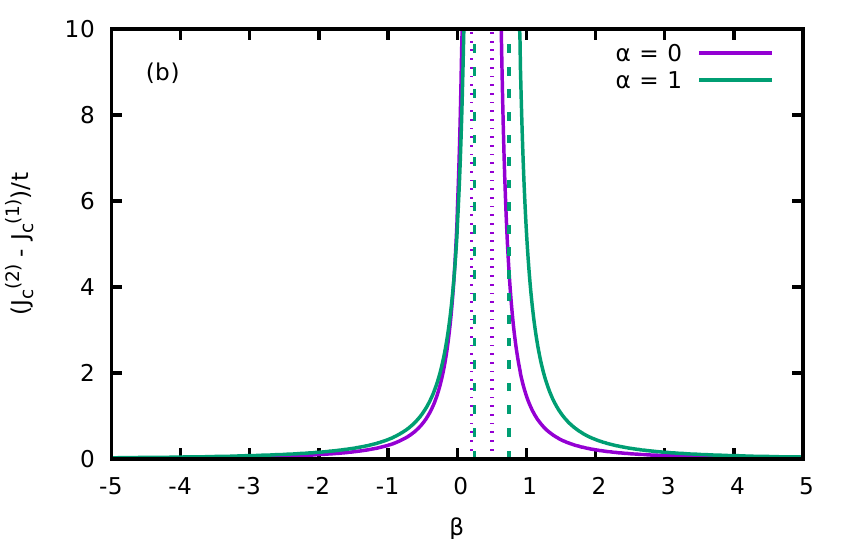}
\includegraphics[width=0.45\textwidth]{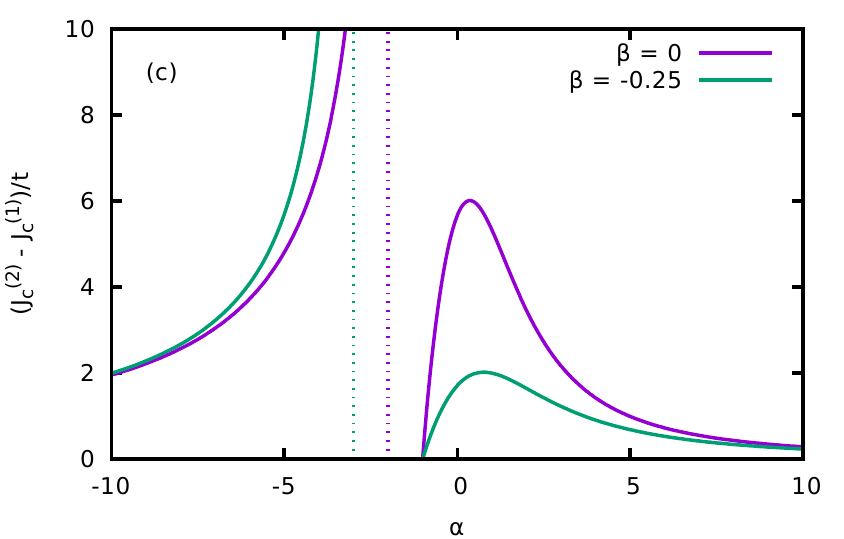}
\caption{(a) Estimate of the size of the superconducting region (in units in which $t \equiv 1$) at low fillings as a function of $\alpha = J_z/J_\perp$ and $\beta = V/ J_\perp$. The result is obtained from the difference of Eqs.~\eqref{eq:Jc2}~minus~\eqref{eq:Jc1-ph}. In the yellow region, the size is greater than or equal to $25$; in the white region, $J_c^{(1)}>J_c^{(2)}$, indicating the absence of SC. (b) and (c) Size of the SC region when keeping $\alpha$ or $\beta$ fixed, as indicated. The vertical dashed and dotted lines show the position of the poles of $\left(J_c^{(2)}-J_c^{(1)} \right)/t$ at which the value becomes negative, indicating the absence of the SC phase.
}
\label{fig:SCregion}
\end{figure}

As $J_z$ and $V$ can be tuned independently from each other in the polar molecule quantum simulators, one can ask for the dependence of the size of this superconducting region as a function of both.
In Fig.~\ref{fig:SCregion} we show the result as obtained from Eqs.~\eqref{eq:Jc2} and~\eqref{eq:Jc1-ph}. 
Note that there is a significant increase, e.g., for $\alpha = 0$ when approaching $\beta = (1 - \sqrt{2})/2$, at which Eq.~\eqref{eq:Jc2} has a divergence.
Also note that there is a region $0.21 \lesssim \beta \lesssim 0.6$ for $\alpha = 0$ in which $J_c^{(2)} - J_c^{(1)}$ is negative, indicating the absence of a superconducting phase.
In this parameter region, phase separation takes over and inhibits the formation of the superconducting phase, which is the precursor region.
In Fig.~\ref{fig:SCregion}(c), the size of the SC region as a function of $\alpha$ when keeping $\beta$ fixed is shown. 
As can be seen, for $\beta =0$ or $\beta = -0.25$ a large increase can also be obtained by tuning $\alpha$ to negative values $-5 \lesssim \alpha \lesssim -3$. 
It therefore appears very promising to study the behavior of the phase diagram in these regions with an enhanced SC phase numerically and in quantum simulators, since the SC phase seems to be maximized there. 

Summarizing the results of this section, we note that both a negative value of $V$ and a positive (antiferromagnetic) value of $J_z$ act as attractive interactions. 
We see that tuning $V$ or $J_z$ to zero leads to the somewhat counterintuitive finding that the superconducting phase gets enhanced when suppressing an attractive interaction.
This is because attractive interaction helps in stabilizing phase separation, if it gets too large, as discussed in Refs.~\onlinecite{Kivelson1990,Dagotto1992,Troyer1993}.
It would be interesting to find the optimal ratios of $V$, $J_z$, and $J_\perp$ for superconductivity.  
It would similarly be interesting to explore the possibility to tune these terms also in materials, in which, e.g., due to spin-orbit couplings, an $XXZ$-type anisotropy in spin-exchange interactions should be possible, and to see how this affects the superconducting properties of such systems.   
   
%%%%%%%%%%%%%%%%%%%%%%%%%%%%%%%%%%%%%%%%%%%%%%%%%%%%%%%%%%%%%%%%%%%%%%%%%%%%%%%%%%%%%%%%%%

\section{Role of dipolar interactions}
\label{sec:longrange}

Power counting shows that interactions decaying  $\sim 1/r^3$ should be irrelevant in 1D~\cite{giamarchi} and hence the phase diagram of a model should not be qualitatively altered when turning on these interactions.
However, we find important qualitative features that this argument fails to capture and in the following we argue that algebraically decaying long-range interactions can significantly alter the behavior of correlation functions.

For example, conventional wisdom holds that correlation functions in gapped phases in any dimension decay exponentially.  
Indeed, Hastings and Koma proved this~\cite{Hastings2006} for general short-range interacting spin systems.  
In contrast, recent theoretical studies have found that long-range interacting systems can have algebraically decaying correlation functions despite the existence of a gap~\cite{DengPRA2005,SchuchCommMathPhys2006,SchachenmayerNJP2010}.
%,peter_anomalous_2012}.
In this section we explore this behavior in our model, where especially interesting features emerge.
The main result is that a full treatment of the long-range interactions is mainly necessary to compute long-distance correlators in the presence of a gap, where one obtains an algebraic tail due to the long-range interactions also in the present case.

\subsection{Relevance of a cutoff in the interaction range} 
\label{subsec:cutoff}

\begin{figure}[t]
\includegraphics{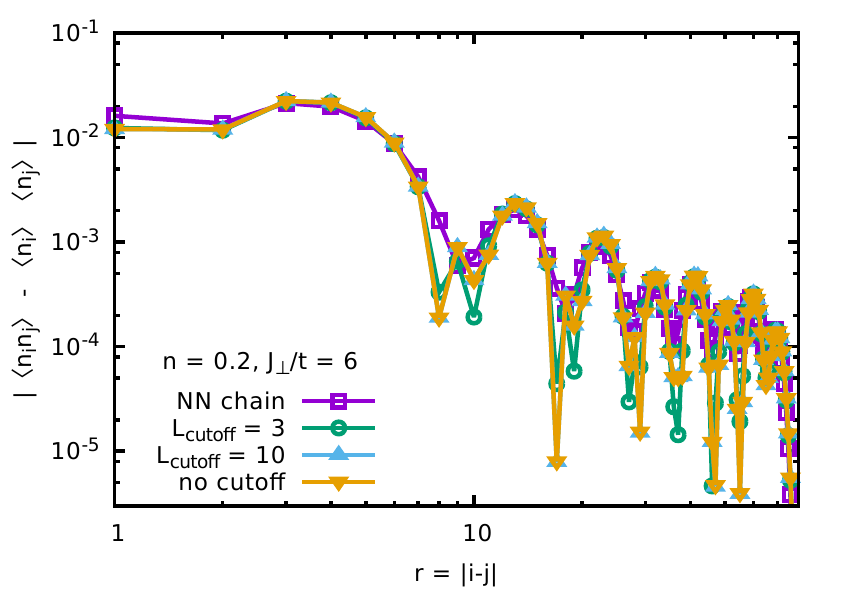}
\caption{Effect of truncating the range of the dipolar interactions on algebraically decaying correlation functions, here the density-density correlation function for $n=0.2$ and $J_\perp/t = 6$. The plot displays results for NN interactions, for a truncation in the interaction range after three and ten sites, and results for the full range of the interactions, as indicated.}
\label{fig:Lcutoffdensdens}
\end{figure}

Numerically, it is a challenge to take into account the interaction terms at all distances.
It is therefore tempting to introduce a cutoff in the long-range character of the interactions.
However, this can lead to wrong results since it can mask the realization of subdominant contributions to the correlation functions and maybe mask further effects. 
Here we analyze to what extent it is necessary to account for long-range interactions in the case of the $t$-$J_\perp$ chain.

In Fig.~\ref{fig:Lcutoffdensdens} we analyze the effect of a cutoff in the interaction range for the density correlations, for which there is no gap, and consequently the correlations are algebraic even for short-range interacting systems.
We compare our results for the density-density correlation function for systems with NN interactions only, with a cutoff in the interactions after three and ten sites and when keeping the full range.
As can be seen, going beyond NN interactions changes quantitatively the behavior of the correlation functions.
The results with a cutoff of three sites, however, are already in qualitative agreement with the ones taking the full range into account and the results with ten sites, for this particular example, are in quantitative agreement with the full-range results, within the estimated accuracy.  
Therefore, as mentioned in Sec.~\ref{subsec:DMRG}, for systems with $L \geq 100$ we use a cutoff in the interaction range of 20 sites, which makes it possible to treat systems with up to 200 lattice sites, while keeping the high accuracy needed to investigate the phase diagram. 

\subsection{Algebraic tail in spin correlation functions} 
\label{subsec:algebraictail}

\begin{figure}[b]
\includegraphics{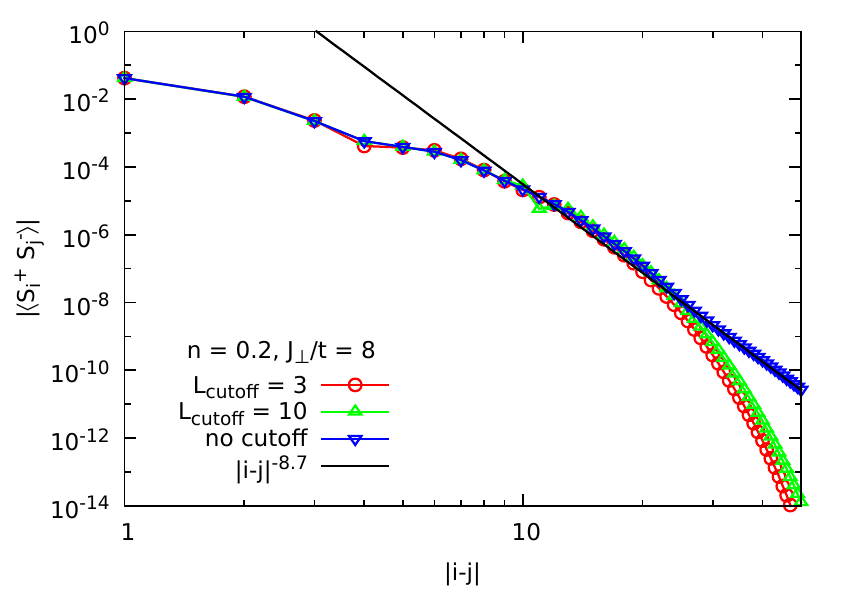}
\caption{Effect of long-range $1/|i-j|^3$ interactions on the transverse spin correlation functions. The results shown are at $n=0.2$ and $J_\perp/t = 8$ for dipolar interactions truncated after three and ten lattices sites and for the full range of interactions, as indicated. The black line is a fit of an algebraic function with exponent $8.7$ in the long-distance part in the case of full-range interactions.}
\label{fig:Lcutoff}
\end{figure}

Here we use the DMRG to show that the spin-spin correlations of the spin-gapped phase of the $t$-$J_\perp$ chain decay algebraically.
We provide a simple analytic framework for understanding this behavior as well as that of other models, such as those in prior theoretical studies~\cite{DengPRA2005,SchuchCommMathPhys2006,SchachenmayerNJP2010}
% SRM: I don't remember which Ref this could be...maybe check later,peter_anomalous_2012}
and including dimensions $d>1$.

Figure~\ref{fig:Lcutoff} compares the DMRG results for $\langle S_i^+ S_j^-\rangle$ obtained using a cutoff of the interaction range at distances of three and ten sites and keeping the full range of interactions.
As can be seen, for the systems with a cutoff the correlations decay exponentially.
However, in the case of full-range interactions, at distances $|i-j|\gtrsim 20$ the behavior is significantly different. A fit (to an admittedly narrow spatial window) indicates a power-law decay in the tail of $\sim |i-j|^\beta$ with $\beta \approx 8.7$.

Now we present a framework to understand the behavior observed numerically. 
We adopt the perspective of imagining starting with a nearest-neighbor interacting model $H_{\text{NN}}$ and turning on long-range interactions $H_{\text{LR}}$ perturbatively.  
Although this is valid only if the long-range interactions are sufficiently small, it should capture the key physics as long as the long-range interactions do not drive the system through a phase transition.  
Fortunately, at least in one-dimensional dipolar chains, the matrix elements of the long-range pieces are suppressed by at least a factor of 8 from the nearest-neighbor case, suggesting that the perturbation theory could frequently be a valid approach. 

Our arguments will be based on perturbation theory in a linear-response formalism, which will allow us to apply some powerful theorems such as that of Lieb and Robinson~\cite{LiebRobinson72}. 
Standard perturbation theory for the difference $\delta \expec{\mc O}$ of some observable $\mc O$  between the nearest-neighbor and nearest-neighbor plus long-range interacting states gives
\begin{equation}
\delta \expec{\mc O} = -i \int_0^\infty \! dt\, \expec{[\mc O(t),H_{\text{LR}}]}_0 ,\label{eq:corrn-from-pert-gen}
\end{equation}
where the expectation value is in the ground state of $H_{\text{NN}}$ and the time evolution of ${\mc O}(t)$ is in the Heisenberg picture of $H_{\text{NN}}$.

Before turning to the $t$-$J_\perp$ model, we will apply our results to simpler cases: first an Ising and then an $XXZ$ model.
These will give us simple constructive examples of gapped phases with algebraic decay. 
It also will highlight a significant difference in the algebraic decay of the $t$-$J_\perp$ chains compared to these other examples, as well as previously studied models~\cite{DengPRA2005,SchuchCommMathPhys2006,SchachenmayerNJP2010}.
%,peter_anomalous_2012}.

First, consider the nearest-neighbor Ising antiferromagnet $J\sum_i S^z_i S^z_{i+1}$ perturbed by long-range transverse interactions $H_{\text{LR}}=(\lambda J/2)\sum_{i,j}\frac{1}{|i-j|^3}\left( S^+_i S^-_{i+1}+\text{H.c.} \right)$ and calculate the change in the observable $\mc O_{ab}= S^+_a S^-_b$ for sites $a$ and $b$ far apart.  
For this case we can straightforwardly solve for $\mc O(t)$ appearing in Eq.~\eqref{eq:corrn-from-pert-gen} and do the integral. 
Using the identity $f(S^z) S^+ = S^+ f(S^z+1)$ and its conjugate, one finds $\expec{[\mc O_{ab}(t),H_{\text{LR}}]}_0 = \lambda e^{iJt(P_a-P_b)} \expec{[\mc O_{ab},H_{\text{LR}}]}_0$ with $P_j =(-1)^j$ in the $H_{\text{NN}}$ ground state $\ket{\cdots\uparrow \downarrow \uparrow \downarrow \cdots}$.
Evaluating the remaining equal time commutator and taking the expectation value, 
one finds $\expec{[\mc O_{ab}(t),H_{\text{LR}}]}_0 = \lambda J P_b e^{iJt(P_a-P_b)} Q_{ab}/|a-b|^3$ where $Q_{ab}$ is the function that is zero if $a$ and $b$ are both even or both odd and unity otherwise.  
Finally, doing the integral,
\begin{equation}
\delta \expec{\mc O_{ab}} = -\lambda\frac{Q_{ab}}{2|a-b|^3}\label{eq:Ising-correlator}
\end{equation}
for well separated $a$ and $b$.

Equation~\eqref{eq:Ising-correlator} shows that perturbing the gapped nearest-neighbor Ising antiferromagnet by dipolar $H_{\text{LR}}$ transverse perturbations gives rise to algebraically decaying transverse correlations.  
Also note that since the long-range interactions are perturbatively small by hypothesis, the gap remains open. 
Thus the phase is adiabatically connected to the phase with exponentially decaying correlations, despite its long-range correlations.

The structure behind this result exists very generally, even in models where we cannot exactly calculate the correlations.  
The integrand of Eq.~\eqref{eq:corrn-from-pert-gen} that determines the response contains $\expec{[\mc O_{ab}(t),H_{\text{LR}}]}_0$, which is a sum of four-operator terms.  
Up to constants, it is $ \sum_{i,j} \frac{1}{|i-j|^3} \expec{[S^+_a(t) S^-_b(t),S^+_i S^-_j + S^+_j S^-_i]}_0$.  
In the Ising model, in order to be nonvanishing, the raising operator at site $a$ needs to pair with a lowering operator and similarly for site $b$'s lowering operator, so the two nonvanishing terms are (i) $a=j$ and $b=i$ and (ii) $a=i$ and $b=j$.  
Consequently, the factor $1/|i-j|^3$ is equal to $1/|a-b|^3$.  

For general models $H_{\text{NN}}$, for example, the $XXZ$ model, the operators such as $S^+_a(t)$ will no longer be localized to a single site $a$, but at least for short times will be localized close to $a$, a consequence of the Lieb-Robinson bound.
There is some subtlety, as we must integrate out to $t=\infty$, and the length scale around site $a$ around which $S_z^+(t)$ is localized grows with time. 
However, under some rather mild assumptions, the integral for large $|a-b|$ is dominated by the regime where the operators are localized compared to the distance between $a$ and $b$. 
In this case, when calculating the correlations on sites $a$ and $b$ we expect a factor of $1/|a-b|^3$ (for large $|a-b|$) coming from $H_{\text{LR}}$ in the correlator. 
Indeed, this agrees with previous numerical findings~\cite{DengPRA2005,SchuchCommMathPhys2006,SchachenmayerNJP2010}:
%,peter_anomalous_2012}:
The power law of the correlation function decay matches the power law of the interaction.

\begin{figure}
\includegraphics{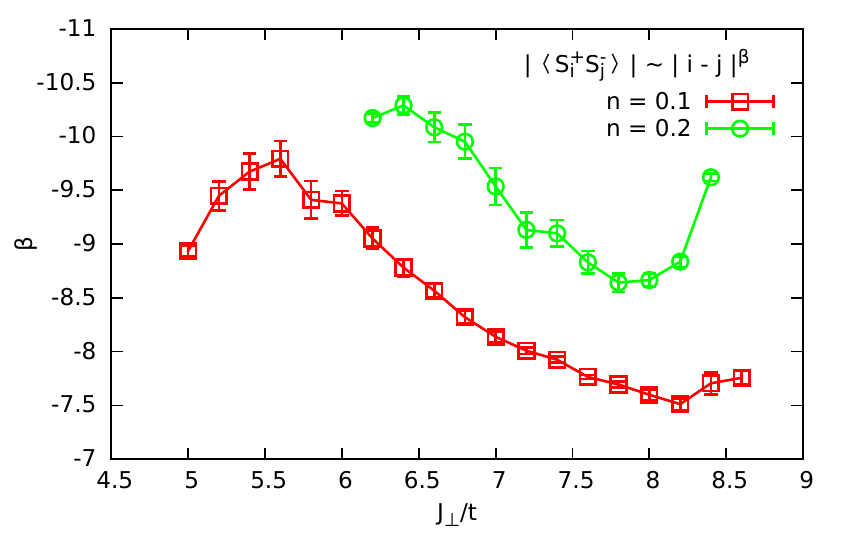}
\caption{Results for the exponent of the algebraic tail for the dipolar $t$-$J_\perp$ chain~\eqref{eq:dipolartJperp} as function of $J_\perp/t$ at fillings $n=0.1$ (red squares) and $n=0.2$ (green circles), obtained from fits to the transverse spin correlation function.}
\label{fig:exponent_tail}
\end{figure}

In light of this analytic result and previous numerical results, the correlations we find in the $t$-$J_\perp$ chain are even more intriguing.  
For example, in Fig.~\ref{fig:Lcutoff}, the correlation function appears to decay (roughly) as $1/r^{8.7}$ despite the $1/r^3$ interaction. 
We have attempted to obtain the exponent of this algebraic tail by fitting to results for $L=80$ sites  and show our results at fillings $n=0.1$ and $n=0.2$ in Fig.~\ref{fig:exponent_tail} (due to the smallness of the systems, the quality of our fits at larger fillings was substantially worse, so we refrain from discussing these cases; for the quality of our fits at low fillings, see Fig.~\ref{fig:Lcutoff}).  
As can be seen, the value of the exponent seems to vary between $\sim 7$ and $\sim 11$.
This is a very wide range and is most probably affected by finite-size effects.
However, it indicates that the value of the exponent can vary with the parameters and is larger than 3, contrary to the previous understanding. 

Based on the considerations developed above, we give a suggestive argument that the decay could be faster than $1/r^3$ for the $t$-$J_\perp$ chain.
For the $t$-$J_\perp$ chain the long-range transverse correlations induced by the long-range transverse dipolar interactions in the $t$-$J_\perp$ chain are $\delta \expec{\mc O_{ab}}=\frac{J_\perp}{2}\sum_{i,j} \frac{1}{|i-j|^3}\int_0^\infty \!dt\, \expec{[S^+_a(t) S^-_b(t),S^+_i S_j^- + S^-_i S_j^+]}_0$.  
This is identical to the Ising example above, except the operator dynamics are under the NN $t$-$J_\perp$ Hamiltonian and the expectation value is in the NN $t$-$J_\perp$ ground state.
Again, $S^+_a(t)$ is localized near $a$. 
However, unlike the Ising case and the $XXZ$ case (in the antiferromagnetic phase) there is no long-range order and hence the expectation value of the commutator vanishes as $|a-b|\rightarrow \infty$ (at the same level of rigor as our earlier arguments).  
This implies that any correlations induced by the long-range interactions must occur as a higher power of the interaction Hamiltonian and will thus decay faster than $1/r^3$. 
In fact, since the expectation value of the correlator will concern operators on two far-separated spatial regions, which in the Luttinger liquid typically decay at least as fast as $1/r^2$, $\delta \expec{{\mc O}_ab}$ likely must decay as $1/r^5$ or faster.
The dominant contribution in the tail could come either from algebraic decay of the correlator in the expectation value or from higher orders of perturbation theory (or both).
It would be interesting to understand the origin of the rapid but nonexponential decay with the exponents found in Figs.~\ref{fig:Lcutoff} and~\ref{fig:exponent_tail}.

\section{Summary}
\label{sec:summary}

In this paper we calculated the phase diagrams of the NN, NNN, and dipolar $t$-$J_\perp$ chains as a function of filling and spin exchange $J_\perp/t$ using the DMRG.
We presented a thorough analysis of correlations that led to the phase diagrams summarized in Fig.~\ref{fig:phasedia}.
Interestingly, turning off the $J_z$ and $V$ term of the original $t$-$J$-model leads to a superconducting phase that occupies more of the phase diagram and possesses a larger spin gap, when measured in units of the hopping $t$.

Adding dipolar long-range interactions changes the phase diagram rather mildly.
The majority of the changes to the boundaries of the phase diagram from dipolar interactions are reproduced already by the NNN interactions. 
However, to reproduce tails of the spin correlation function in the spin-gapped phase it is necessary to retain the long-range interactions. 

All of these behaviors were reproduced qualitatively with analytic arguments that we gave in Secs.~\ref{sec:variational} and~\ref{sec:longrange}. 
Simple variational \textit{Ans\"atze} for each phase allow us to understand the phase diagram for the $t$-$J_\perp$ model considered in this paper and predict the behavior of others with $J_z\ne 0$ and $V\ne 0$. 
Our analytic arguments extend straightforwardly to higher dimensions and may guide experiments with ultracold molecules as well as those searching for robust superconductors. 
Understanding the behavior for finite $J_z$ and $V$ and especially in higher dimensions quantitatively is an interesting future challenge. 

\acknowledgements
We acknowledge useful discussions with A. V. Gorshkov and A. M. Rey. 
We thank A. Moreno for providing the results for the standard $t$-$J$ chain in Fig.~\ref{fig:spingap_n0.1}, which in part were presented in Ref.~\onlinecite{Moreno} (co-authored by S.R.M.). 
We acknowledge financial support from PIF-NSF (Grant No. 0904017).
K.R.A.H. was supported in  part with  funds from the Welch Foundation (Grant No.  C-1872). 
M.M. gratefully acknowledges support from a fellowship of the Studienstiftung des Deutschen Volkes e.V.
M.M. and S.R.M. acknowledge financial support from the Deutsche Forschungsgemeinschaft (DFG) through SFB/CRC 1073 (Projects No. A05 and No. B03).
S.R.M. and K.R.A.H. acknowledge the Kavli Institute for Theoretical Physics (KITP) and K.R.A.H. acknowledges the Aspen Center for Physics where part of this research was accomplished and supported in part by the NSF under Grants No. NSF PHY11-25915 and No. PHY-1607611. 
%This work was partially performed at the Aspen Center for Physics, which is supported by National Science Foundation grant PHY-1607611.
This work utilized the Janus supercomputer, which is supported by the NSF (Award No. CNS-0821794) and CU Boulder. The Janus supercomputer is a joint effort of the University of Colorado Boulder, the University of Colorado Denver, and the National Center for Atmospheric Research.

\bibliographystyle{prsty}
\bibliography{biblio}

\begin{thebibliography}{10}

\bibitem{reviewmolecules}
L.~D. Carr, D. DeMille, R.~V. Krems, and J. Ye, New Journal of Physics {\bf
  11},  055049  (2009).

\bibitem{Lemeshko_Review}
M. Lemeshko, R.~V. Krems, J.~M. Doyle, and S. Kais, Molecular Physics {\bf
  111},  1648  (2013).

\bibitem{PhysRevLett.101.133004}
J. Deiglmayr, A. Grochola, M. Repp, K. M\"ortlbauer, C. Gl\"uck, J. Lange, O.
  Dulieu, R. Wester, and M. Weidem\"uller, Phys. Rev. Lett. {\bf 101},  133004
  (2008).

\bibitem{Silke_science}
K.-K. Ni, S. Ospelkaus, M.~H.~G. de~Miranda, A. Pe'er, B. Neyenhuis, J.~J.
  Zirbel, S. Kotochigova, P.~S. Julienne, D.~S. Jin, and J. Ye, Science {\bf
  322},  231  (2008).

\bibitem{Silke_science2}
S. Ospelkaus, K.-K. Ni, D. Wang, M.~H.~G. de~Miranda, B. Neyenhuis, G.
  Qu\'em\'ener, P.~S. Julienne, J.~L. Bohn, D.~S. Jin, and J. Ye, Science {\bf
  327},  853  (2010).

\bibitem{Silke_Nature}
K.~K. Ni, S. Ospelkaus, D. Wang, G. Quemener, B. Neyenhuis, M.~H.~G.
  de~Miranda, J.~L. Bohn, J. Ye, and D.~S. Jin, Nature {\bf 464},  1324
  (2010).

\bibitem{PhysRevLett.104.030402}
S. Ospelkaus, K.-K. Ni, G. Qu\'em\'ener, B. Neyenhuis, D. Wang, M.~H.~G.
  de~Miranda, J.~L. Bohn, J. Ye, and D.~S. Jin, Phys. Rev. Lett. {\bf 104},
  030402  (2010).

\bibitem{Amodsen}
A. Chotia, B. Neyenhuis, S.~A. Moses, B. Yan, J.~P. Covey, M. Foss-Feig, A.~M.
  Rey, D.~S. Jin, and J. Ye, Phys. Rev. Lett. {\bf 108},  080405  (2012).

\bibitem{PhysRevLett.84.246}
A.~N. Nikolov, J.~R. Ensher, E.~E. Eyler, H. Wang, W.~C. Stwalley, and P.~L.
  Gould, Phys. Rev. Lett. {\bf 84},  246  (2000).

\bibitem{PhysRevLett.94.203001}
J.~M. Sage, S. Sainis, T. Bergeman, and D. DeMille, Phys. Rev. Lett. {\bf 94},
  203001  (2005).

\bibitem{PhysRevLett.101.133005}
F. Lang, K. Winkler, C. Strauss, R. Grimm, and J.~H. Denschlag, Phys. Rev.
  Lett. {\bf 101},  133005  (2008).

\bibitem{deMiranda:2011gd}
M.~H.~G. de~Miranda, A. Chotia, B. Neyenhuis, D. Wang, G. Qu{\'e}m{\'e}ner, S.
  Ospelkaus, J.~L. Bohn, J. Ye, and D.~S. Jin, Nature Physics {\bf 7},  502
  (2011).

\bibitem{PhysRevLett.112.070404}
B. Zhu, B. Gadway, M. Foss-Feig, J. Schachenmayer, M.~L. Wall, K.~R.~A.
  Hazzard, B. Yan, S.~A. Moses, J.~P. Covey, D.~S. Jin, J. Ye, M. Holland, and
  A.~M. Rey, Phys. Rev. Lett. {\bf 112},  070404  (2014).

\bibitem{focus_ultracoldmolecules}
L.~D. Carr and J. Ye, New Journal of Physics {\bf 11},  055009  (2009).

\bibitem{PhysRevLett.114.205302}
J.~W. Park, S.~A. Will, and M.~W. Zwierlein, Phys. Rev. Lett. {\bf 114},
  205302  (2015).

\bibitem{PhysRevLett.116.225306}
S.~A. Will, J.~W. Park, Z.~Z. Yan, H. Loh, and M.~W. Zwierlein, Phys. Rev.
  Lett. {\bf 116},  225306  (2016).

\bibitem{review_molecules2}
B. Gadway and B. Yan, Journal of Physics B: Atomic, Molecular and Optical
  Physics {\bf 49},  152002  (2016).

\bibitem{MolonyPRL}
P.~K. Molony, P.~D. Gregory, Z. Ji, B. Lu, M.~P. K\"oppinger, C.~R. Le~Sueur,
  C.~L. Blackley, J.~M. Hutson, and S.~L. Cornish, Phys. Rev. Lett. {\bf 113},
  255301  (2014).

\bibitem{TakekoshiPRL}
T. Takekoshi, L. Reichs\"ollner, A. Schindewolf, J.~M. Hutson, C.~R. Le~Sueur,
  O. Dulieu, F. Ferlaino, R. Grimm, and H.-C. N\"agerl, Phys. Rev. Lett. {\bf
  113},  205301  (2014).

\bibitem{GoulvenPRL2016}
M. Guo, B. Zhu, B. Lu, X. Ye, F. Wang, R. Vexiau, N. Bouloufa-Maafa, G.
  Qu\'em\'ener, O. Dulieu, and D. Wang, Phys. Rev. Lett. {\bf 116},  205303
  (2016).

\bibitem{Manin}
Y.~I. Manin, Computable and uncomputable (in Russian); Moscow, Sovetskoye Radio
  1980. See appendix of arXiv:quant-ph/9903008 for an english translation.

\bibitem{feynman1}
R.~P. Feynman, Int. J. Theor. Phys. {\bf 21},  467  (1982).

\bibitem{feynman2}
R.~P. Feynman, Found. Phys. {\bf 16},  507  (1985).

\bibitem{feynman3}
R.~P. Feynman, Optics News {\bf 11},  11  (February 1985).

\bibitem{MolPhys2013}
A.~V. Gorshkov, K.~R. Hazzard, and A.~M. Rey, Molecular Physics {\bf 111},
  1908  (2013).

\bibitem{micheli_naturephysics}
A. Micheli, G.~K. Brennen, and P. Zoller, Nature Physics {\bf 2},  341
  (2006).

\bibitem{pupillo_polargases}
G. Pupillo, A. Micheli, H.~P. B\"uchler, and P. Zoller,  in {\em Cold
  Molecules: Theory, Experiment, Applications}, edited by R. Krems, W.
  Stwallye, and B. Friedrich (CRC Press, Boca Raton, FL, 2009).

\bibitem{PhysRevB.87.081106}
S.~R. Manmana, E.~M. Stoudenmire, K.~R.~A. Hazzard, A.~M. Rey, and A.~V.
  Gorshkov, Phys. Rev. B {\bf 87},  081106  (2013).

\bibitem{KadenPRL2014}
K.~R.~A. Hazzard, B. Gadway, M. Foss-Feig, B. Yan, S.~A. Moses, J.~P. Covey,
  N.~Y. Yao, M.~D. Lukin, J. Ye, D.~S. Jin, and A.~M. Rey, Phys. Rev. Lett.
  {\bf 113},  195302  (2014).

\bibitem{springer_quantummagnetism}
{\em Quantum Magnetism}, Vol.~645 of {\em Lecture Notes in Physics}, edited by
  U. Schollw\"{o}ck, J. Richter, D. Farnell, and R. Bishop (Springer,
  Berlin/Heidelberg, 2004).

\bibitem{book_frustratedspins}
{\em Frustrated Spin Systems}, edited by H.~T. Diep (World Scientific
  Publishing, Singapore, 2004).

\bibitem{book_HFMTrieste}
{\em Introduction to Frustrated Magnetism}, Vol.~164 of {\em Springer Series in
  Solid-State Sciences}, edited by C. Lacroix, P. Mendels, and F. Mila
  (Springer, Berlin / Heidelberg, 2011).

\bibitem{Simon:2011p2830}
J. Simon, W.~S. Bakr, R. Ma, M. Tai, P. Preiss, and M. Greiner, Nature {\bf
  472},  307  (2011).

\bibitem{1367-2630-9-5-138}
G.~K. Brennen, A. Micheli, and P. Zoller, New Journal of Physics {\bf 9},  138
  (2007).

\bibitem{reviewmolecules_quantummagnetism}
M.~L. Wall, K.~R.~A. Hazzard, and A.~M. Rey, arXiv:1406.4758  (2014).

\bibitem{PhysRevLett.110.075301}
K.~R.~A. Hazzard, S.~R. Manmana, M. Foss-Feig, and A.~M. Rey, Phys. Rev. Lett.
  {\bf 110},  075301  (2013).

\bibitem{Yan:2013fn}
B. Yan, S.~A. Moses, B. Gadway, J.~P. Covey, K.~R.~A. Hazzard, A.~M. Rey, D.~S.
  Jin, and J. Ye, Nature {\bf 501},  521  (2013).

\bibitem{PhysRevA.84.033619}
A.~V. Gorshkov, S.~R. Manmana, G. Chen, E. Demler, M.~D. Lukin, and A.~M. Rey,
  Phys. Rev. A {\bf 84},  033619  (2011).

\bibitem{dagotto}
E. Dagotto, Rev. Mod. Phys. {\bf 66},  763  (1994).

\bibitem{PhysRevLett.107.115301}
A.~V. Gorshkov, S.~R. Manmana, G. Chen, J. Ye, E. Demler, M.~D. Lukin, and
  A.~M. Rey, Phys. Rev. Lett. {\bf 107},  115301  (2011).

\bibitem{Bloch:2005p988}
I. Bloch, Nature Physics {\bf 1},  23  (2005).

\bibitem{Bloch:2008p943}
I. Bloch, J. Dalibard, and W. Zwerger, Rev. Mod. Phys. {\bf 80},  885  (2008).

\bibitem{tJ1977}
K.~A. Chao, J. Spalek, and A.~M. Oles, Journal of Physics C: Solid State
  Physics {\bf 10},  L271  (1977).

\bibitem{tJoriginal1}
P.~W. Anderson, Science {\bf 235},  1196  (1987).

\bibitem{tJoriginal2}
F.~C. Zhang and T.~M. Rice, Phys. Rev. B {\bf 37},  3759  (1988).

\bibitem{auerbach}
A. Auerbach, {\em Interacting Electrons and Quantum Magnetism} (Springer, New
  York, 1994).

\bibitem{MaylePRA2012}
M. Mayle, B.~P. Ruzic, and J.~L. Bohn, Phys. Rev. A {\bf 85},  062712  (2012).

\bibitem{MaylePRA2013}
M. Mayle, G. Qu\'em\'ener, B.~P. Ruzic, and J.~L. Bohn, Phys. Rev. A {\bf 87},
  012709  (2013).

\bibitem{AndrisPRL2016}
A. Do\ifmmode~\mbox{\c{c}}\else \c{c}\fi{}aj, M.~L. Wall, R. Mukherjee, and
  K.~R.~A. Hazzard, Phys. Rev. Lett. {\bf 116},  135301  (2016).

\bibitem{MichaelWallPRA2017}
M.~L. Wall, N.~P. Mehta, R. Mukherjee, S.~S. Alam, and K.~R.~A. Hazzard, Phys.
  Rev. A {\bf 95},  043635  (2017).

\bibitem{MichaelWallPRA2017_2}
M.~L. Wall, R. Mukherjee, S.~S. Alam, N.~P. Mehta, and K.~R.~A. Hazzard, Phys.
  Rev. A {\bf 95},  043636  (2017).

\bibitem{white1992}
S.~R. White, Phys. Rev. Lett. {\bf 69},  2863  (1992).

\bibitem{white1993}
S.~R. White, Phys. Rev. B {\bf 48},  10345  (1993).

\bibitem{dmrgbook}
{\em Density Matrix Renormalization - A New Numerical Method in Physics},
  edited by I. Peschel, X. Wang, M. Kaulke, and K. Hallberg (Springer Verlag,
  Berlin, 1999).

\bibitem{noack:93}
R.~M. Noack and S.~R. Manmana, AIP Conference Proceedings {\bf 789},  93
  (2005).

\bibitem{Schollwock:2005p2117}
U. Schollw{\"o}ck, Rev. Mod. Phys. {\bf 77},  259  (2005).

\bibitem{Schollwock:2011p2122}
U. Schollw{\"o}ck, Annals of Physics {\bf 326},  96  (2011).

\bibitem{Moreno}
A. Moreno, A. Muramatsu, and S.~R. Manmana, Phys. Rev. B {\bf 83},  205113
  (2011).

\bibitem{giamarchi}
T. Giamarchi, {\em Quantum Physics in One Dimension}, Vol.~121 of {\em
  International Series of Monographs on Physics} (Oxford University Press,
  Oxford, 2004).

\bibitem{Troyer1993}
M. Troyer, H. Tsunetsugu, T.~M. Rice, J. Riera, and E. Dagotto, Phys. Rev. B
  {\bf 48},  4002  (1993).

\bibitem{DengPRA2005}
X.-L. Deng, D. Porras, and J.~I. Cirac, Phys. Rev. A {\bf 72},  063407  (2005).

\bibitem{SchachenmayerNJP2010}
J. Schachenmayer, I. Lesanovsky, A. Micheli, and A.~J. Daley, New Journal of
  Physics {\bf 12},  103044  (2010).

\bibitem{AlexeyGuidoPRL2014}
D. Vodola, L. Lepori, E. Ercolessi, A.~V. Gorshkov, and G. Pupillo, Phys. Rev.
  Lett. {\bf 113},  156402  (2014).

\bibitem{SchuchCommMathPhys2006}
N. Schuch, J.~I. Cirac, and M.~M. Wolf, Communications in Mathematical Physics
  {\bf 267},  65  (2006).

\bibitem{voitreview}
J. Voit, Rep. Prog. Phys. {\bf 58},  977  (1995).

\bibitem{BatistaOrtizPRLtJz}
C.~D. Batista and G. Ortiz, Phys. Rev. Lett. {\bf 85},  4755  (2000).

\bibitem{book_hubbardmodel}
F.~H.~L. Essler, H. Frahm, F. G\"ohmann, A. Kl\"umper, and V.~E. Korepin, {\em
  The One-Dimensional Hubbard Model} (Cambridge University Press, Cambridge,
  2005).

\bibitem{Voit}
J. Voit, J. Phys. C: Solid State Phys. {\bf 21},  L1141  (1988).

\bibitem{ejima05}
S. Ejima, F. Gebhard, and S. Nishimoto, Europhys. Lett. {\bf 70},  492  (2005).

\bibitem{pruschke92}
T. Pruschke and H. Shiba, Phys. Rev. B {\bf 46},  356  (1992).

\bibitem{BLBQ}
S.~R. Manmana, A.~M. L\"auchli, F.~H.~L. Essler, and F. Mila, Phys. Rev. B {\bf
  83},  184433  (2011).

\bibitem{CRAN_GA}
{CRAN~R~project}, GA: Genetic algorithms, 2016.

\bibitem{PaeckelSciPost}
S. {Paeckel}, T. {K{\"o}hler}, and S.~R. {Manmana}, arXiv:1706.05338  (2017).

\bibitem{MichaudPRB2010}
F. Michaud, T. Coletta, S.~R. Manmana, J.-D. Picon, and F. Mila, Phys. Rev. B
  {\bf 81},  014407  (2010).

\bibitem{Dagotto1992}
E. Dagotto and J. Riera, Phys. Rev. B {\bf 46},  12084  (1992).

\bibitem{Kivelson1990}
S.~A. Kivelson, V.~J. Emery, and H.~Q. Lin, Phys. Rev. B {\bf 42},  6523
  (1990).

\bibitem{Hastings2006}
M.~B. Hastings and T. Koma, Communications in Mathematical Physics {\bf 265},
  781  (2006).

\bibitem{LiebRobinson72}
E.~H. Lieb and D.~W. Robinson, Commun. Math. Phys. {\bf 28},  251  (1972).

\end{thebibliography}

\end{document}